\documentclass[12pt, letterpaper]{article}
\usepackage{url,color,mathptmx,amsmath,amsthm,amsfonts}
\usepackage[margin=1in]{geometry}
\usepackage[sort&compress,super,comma]{natbib}
\usepackage{graphicx,caption}

\newcommand{\btheta}{\boldsymbol{\theta}}
\DeclareMathOperator*{\argmaxA}{arg\,max}

\newtheorem{theorem}{Theorem}

\begin{document}

\title{Developing Biomarker Combinations in Multicenter Studies via Direct Maximization and Penalization}

\author{Allison Meisner\\[2pt]
\textit{Department of Biostatistics, Johns Hopkins University, Baltimore, MD, USA}\\[2pt]
{Corresponding author: ameisne1@jhu.edu}\\[4pt]
Chirag R. Parikh\\[2pt]
\textit{Division of Nephrology, Johns Hopkins University, Baltimore, MD, USA} \\[4pt]
Kathleen F. Kerr \\[2pt]
\textit{Department of Biostatistics, University of Washington, Seattle, WA, USA}}

\date{}

\maketitle

\begin{abstract}
Motivated by a study of acute kidney injury, we consider the setting of biomarker studies involving patients at multiple centers where the goal is to develop a biomarker combination for diagnosis, prognosis, or screening. As biomarker studies become larger, this type of data structure will be encountered more frequently. In the presence of multiple centers, one way to assess the predictive capacity of a given combination is to consider the center-adjusted AUC (aAUC), a summary of the ability of the combination to discriminate between cases and controls in each center. Rather than using a general method, such as logistic regression, to construct the biomarker combination, we propose directly maximizing the aAUC. Furthermore, it may be desirable to have a biomarker combination with similar performance across centers. To that end, we allow for penalization of the variability in the center-specific AUCs. We demonstrate desirable asymptotic properties of the resulting combinations. Simulations provide small-sample evidence that maximizing the aAUC can lead to combinations with improved performance. We also use simulated data to illustrate the utility of constructing combinations by maximizing the aAUC while penalizing variability. Finally, we apply these methods to data from the study of acute kidney injury.


\textbf{Short title:} Biomarker combinations via maximization and penalization
\end{abstract}

\section{Introduction}

Multicenter studies have long been used in therapeutic settings as a way to increase power and improve generalizability and are increasingly common in biomarker studies~(e.g., \cite{feldstein2009,degos2010,nickolas2012}). In addition, it is now feasible to measure many biomarkers on each participant. As the performance of individual biomarkers is often modest, there is interest in developing combinations of biomarkers for prognosis, diagnosis, and screening. When studies of multiple biomarkers also involve multiple centers, the central question becomes how such biomarker combinations should be constructed. In a given multicenter study, only a sample of centers are observed. Without making strong distributional assumptions, it is not possible to provide predicted probabilities for individuals in a new center. Acknowledging this inherent limitation of the study design, we aim to identify promising biomarker combinations for further development. 

The Translational Research Investigating Biomarker Endpoints in Acute Kidney Injury (TRIBE-AKI) study is an example of a multicenter biomarker study. The TRIBE-AKI study collected data from 1219 cardiac surgery patients at six centers in North America~\cite{parikh2011}. Study participants were followed for diagnosis of acute kidney injury (AKI) during hospitalization; in other words, the participants were hospitalized in order to undergo cardiac surgery and were followed postoperatively for signs of AKI. For each patient, blood and urine were collected at multiple time points pre- and postoperatively, and several biomarkers were measured at each time point. AKI is typically diagnosed via changes in serum creatinine, but these changes often do not happen until several days after the injury occurs. The goal of the study was to identify a combination of biomarkers that can provide an earlier diagnosis of AKI.

Methods to construct biomarker combinations by maximizing the area under the receiver operating characteristic (ROC) curve (AUC) have been proposed. However, in the multicenter setting, there is interest in conditional performance. One measure of conditional performance is the center-adjusted AUC (aAUC). We propose a method to construct linear biomarker combinations by targeting the aAUC. We then extend our method to allow for penalization of the variability in center-specific performance, yielding combinations with good overall performance and more similar performance across centers. We provide theoretical and empirical justification for the proposed methods and apply these methods to data from the TRIBE-AKI study. 

\section{Background}  

Let $D$ be a binary outcome, where cases are denoted by $D=1$ or the subscript $D$ and controls are denoted by $D=0$ or the subscript $\bar{D}$. 

\subsection{Center-adjusted AUC}

Without loss of generality, suppose that for a given predictor $Z$, higher values of $Z$ are more indicative of $D = 1$. Thus, for a particular threshold $\delta$, the true and false positive rates are $P(Z > \delta|D=1)$ and $P(Z > \delta|D=0)$, respectively. The ROC curve for $Z$ plots the true positive rate versus the false positive rate over the range of possible thresholds $\delta$~\cite{pepebook}. The area under the ROC curve (AUC) is a measure of the ability of $Z$ to discriminate between cases and controls, one aspect of the predictive capacity of $Z$. The ROC curve for a useless predictor lies on the 45-degree line, and the corresponding AUC is 0.5~\cite{pepebook}. The ROC curve for a perfect predictor reaches the upper left-hand corner of the unit square, and its AUC is 1~\cite{pepebook}. The AUC can also be interpreted as the probability that $Z$ for a randomly chosen case is larger than $Z$ for a randomly chosen control~\cite{pepebook}. 

In the multicenter setting, $Z$ can be evaluated marginally, by considering the AUC for $Z$ pooled across centers, or conditionally, by summarizing center-specific AUCs. If a measure of marginal performance (i.e., the AUC for $Z$ pooled across centers) is used, center is allowed to potentially influence the assessment of the discriminatory ability of $Z$, restricting interpretability and generalizability~\cite{janes2008}. Instead, performance should be assessed within each center and then summarized across centers; this is analogous to the center-adjusted odds ratio in the etiologic setting and the center-adjusted treatment effect in the therapeutic setting~\cite{kahan2014,janes2008}. One such summary measure is the center-adjusted AUC (aAUC).

The center-adjusted ROC ($aROC_Z$) and corresponding center-adjusted AUC ($aAUC_Z$) of $Z$, proposed by \cite{janes2009b}, can be written as
\begin{align*}
aAUC_Z & = \int_0^1 aROC_Z(t)dt = \int_0^1 \sum_c ROC_{Z|C=c}(t) P(C=c|D=1) dt = \sum_c w_c AUC_{Z|C=c}, 
\end{align*}
where $C$ indicates center, $t$ denotes the false positive rate, $ROC_{Z|C=c}$ and $AUC_{Z|C=c}$ denote the center-specific ROC and AUC, respectively, in center $c$, and $w_c = P(C=c|D=1)$ is the proportion of cases in center $c$. The aAUC is a weighted average of the center-specific AUCs~\cite*{janes2009}. Thus, it is a summary of the ability of $Z$ to discriminate between cases and controls in each center~\cite{janes2008}. When a sample of centers is used to estimate the aAUC of $Z$, the estimate provides insight into the performance of $Z$ in new centers, to the extent that the new centers are similar to those used to estimate the aAUC. This is analogous to applying an estimate of a center-adjusted treatment effect from a multicenter randomized trial to a new center. 

\subsection{Biomarker combinations}

For a collection of biomarkers $\textbf{X}$, the combination $P(D=1|\textbf{X})$ (and monotone increasing functions thereof) is optimal in terms of maximizing the true positive rate at each false positive rate~\cite{mcintosh2002}. Thus, to the extent that the linear logistic model holds, that is, $P(D=1|\textbf{X}) = \mbox{expit}(\btheta^{\top}\textbf{X})$ for some coefficient vector $\btheta$, the combination $\btheta^{\top}\textbf{X}$ is optimal. As the linear logistic model may not hold, methods have been developed to construct biomarker combinations by maximizing the AUC without relying on this model~\cite*{pepe2006}.

Methods have also been developed to identify combinations of biomarkers that maximize the AUC while accommodating covariates~\cite*{liu2013,schisterman2004}. However, implementation of the method proposed by~\cite{liu2013} is computationally challenging or prohibitive for more than two biomarkers. The method proposed by~\cite{schisterman2004} assumes that the biomarkers have multivariate normal distributions and requires specification of the relationship between the covariates (e.g., center) and the biomarkers. Consequently, these methods may not be broadly applicable.  

If the same data are used to construct a biomarker combination and evaluate its performance (by estimating the aAUC, for example), the resulting estimate of performance is optimistically biased~\cite{copas2002}. This bias, which we refer to as ``resubstitution bias"~\cite{kerr2015}, can be addressed by using a bootstrapping procedure to estimate the bias and correct the apparent estimate of performance~\cite{copas2002,harrell2001regression}. Bootstrapping assumes the observations are exchangeable, but in a multicenter study, observations from the same center may be more similar than observations from different centers, leading to clustering of the data by center; thus, bootstrap resampling by center has been suggested~\cite{bouwmeester2013AJE,vanoirbeek2010,localio2001,janes2009}. However, similar results for the center-specific AUC have been found whether resampling was done on centers or individual observations~\cite{bouwmeester2013AJE}. 

\subsection{Smooth AUC approximations}

When logistic regression is used to construct a combination, the fitted combination is obtained by maximizing the logistic likelihood. However, we are interested in using combinations for diagnosis, prognosis, or screening. This motivates maximizing measures of performance, e.g., the AUC, to construct a combination. Such direct maximization is compelling as it matches the objective function to the intended use of the combination~\cite{pepe2000,pepe2006}. One benefit of directly maximizing the AUC is that the resulting combination is optimal, in terms of the AUC, within the class of combinations considered~\cite{pepe2006}. Furthermore, the AUC of a linear combination constructed by targeting the AUC will be at least as large as the AUC for the individual biomarkers~\cite{pepe2000}. This simple, desirable property might not hold when a combination is constructed by maximizing a likelihood.

In practice, for a set of biomarkers $\textbf{X}$, the true AUC for the linear combination $\btheta^\top \textbf{X}$ is unknown. Instead, we can consider the empirical AUC, 
\begin{equation*}
\hat{AUC}(\btheta) = \frac{1}{n_D n_{\bar{D}} } \sum_{i=1}^{n_D} \sum_{j=1}^{n_{\bar{D}}} 1(\btheta^\top \textbf{X}_{Di} > \btheta^\top \textbf{X}_{\bar{D}j}),
\end{equation*}
where $n_D$ and $n_{\bar{D}}$ are the number of cases and controls, respectively, $1(\cdot)$ is the indicator function, $\textbf{X}_{Di}$ denotes the biomarker vector for the $i^{th}$ case, and $\textbf{X}_{\bar{D}j}$ denotes the biomarker vector for the $j^{th}$ control. Since $\hat{AUC}$ involves indicator functions, direct maximization is challenging. However, smooth approximations to the empirical AUC have been proposed.~\cite{lin2011} used a smooth approximation based on the the probit function:
\begin{align*}
\hat{AUC}(\btheta) & \approx R_n(\btheta) = \frac{1}{n_D n_{\bar{D}}} \sum_{i=1}^{n_D} \sum_{j=1}^{n_{\bar{D}}} \Phi\left\lbrace \btheta^{\top} (\textbf{X}_{Di} - \textbf{X}_{\bar{D}j})/h\right\rbrace,
\end{align*}
where $\Phi$ is the standard normal distribution function and $h$ is a tuning parameter. The function $\Phi(v/h)$ serves as an approximation to the indicator function $I(v > 0)$ and $h$ represents the trade-off between approximation accuracy and estimation feasibility and tends to zero as the sample size grows~\cite{lin2011}. Note that $R_n$ is not convex. Other approximations have been proposed, including the logistic function~\cite{ma2007} and the convex ramp function~\cite{fong2016}. The probit function approximation tends to be more accurate and stable than the logistic function approximation~\cite{lin2011} and is more straightforward to implement than the ramp function approximation. 

\section{Methods}

We consider a population of $M$ centers where center $c$ has $N_c$ observations, $c = 1,...,M$. We observe data from $m$ centers with $n_c$ observations from center $c$, giving $n$ total observations. We consider a $p$-dimensional vector of biomarkers $\textbf{X}$. Let $(\textbf{X}, D)$ be the biomarkers and outcome for an arbitrary observation. We use the subscript $i$ on $\textbf{X}$ and $D$ to denote the biomarkers and outcome, respectively, for the $i^{th}$ observation. We use the superscript $c$ on $\textbf{X}$ and $D$ to denote the biomarkers and outcome, respectively, for an observation from center $c$ and we use $n_D^c$ to indicate the number of cases in center $c$. We assume the center-specific disease prevalence is non-trivial, i.e., $P(D=1|C=c) \in (0,1)$, $c = 1,...,M$. We consider linear biomarker combinations, $\btheta^\top \textbf{X}$, as they are often a reasonable choice and have intuitive appeal for clinical collaborators. 

\subsection{Direct maximization}

We are interested in the aAUC for a linear combination of biomarkers defined by $\btheta$, $aAUC(\btheta) = \sum_{c=1}^M w_c AUC_c(\btheta),$ where $w_c = P(C=c|D=1)$ and $AUC_c(\btheta) = P(\btheta^{\top}\textbf{X}^{c}_D > \btheta^{\top}\textbf{X}^{c}_{\bar{D}})$. As with the unadjusted AUC, in practice the aAUC is unknown and we instead consider the empirical aAUC. The empirical aAUC, $\hat{aAUC} = \sum_{c=1}^m \hat{w}_c \hat{AUC}_c(\btheta)$, is based on empirical estimates of the weights, $\hat{w}_c = \frac{n_D^c}{n_D}$, and the center-specific AUCs, $\hat{AUC}_c(\btheta) = \frac{1}{n^c_{D}n^c_{\bar{D}}} \sum_{i=1}^{n_D^c} \sum_{j=1}^{n_{\bar{D}}^c} 1(\btheta^{\top}\textbf{X}_{Di}^{c} > \btheta^{\top}\textbf{X}_{\bar{D}j}^{c}).$ Again, $\hat{aAUC}$ is a function of $\hat{AUC}_c$, which involves indicator functions, making direct maximization challenging. However, we can use a smooth approximation to $\hat{AUC}_c$, which in turn provides a smooth approximation to $\hat{aAUC}$. In particular, we propose the SaAUC estimate
\begin{equation} 
\hat{\btheta} = \argmaxA_{\btheta \in \boldsymbol{\Theta}} aR_n(\btheta), \label{maxaAUCbeta}
\end{equation}
where $\boldsymbol{\Theta} = \lbrace \btheta \in \mathbb{R}^p: ||\btheta|| = 1 \rbrace$, 
\begin{align*}
R^c_{n_c}(\btheta) & = \frac{1}{n^c_D n^c_{\bar{D}}} \sum_{i=1}^{n_D^c} \sum_{j=1}^{n_{\bar{D}}^c} \Phi\left\lbrace \btheta^{\top} (\textbf{X}^c_{Di} - \textbf{X}^c_{\bar{D}j})/h_c\right\rbrace \approx \hat{AUC}_c(\btheta).
\end{align*}
In this approximation, $h_c$ is a tuning parameter that tends to zero as $n_c$ grows. In order to retain identifiability, constraints must be imposed on $\btheta$; in particular, for any given scalar $\kappa$, $aR_n(\btheta) = aR_n(\kappa\btheta)$. Thus, we constrain $||\btheta||=1$ as in \cite{fong2016}. 

The objective function (\ref{maxaAUCbeta}) is a sum of smooth functions and is therefore also smooth. Consequently, gradient-based methods (with Lagrange multipliers to incorporate the $||\btheta||=1$ constraint) can be used to obtain $\hat{\btheta}$, though this procedure may not yield a global maximum since $aR_n(\btheta)$ is not convex. Gradient-based methods require starting values $\tilde{\btheta}$; one possibility is to use estimates from logistic regression. Additionally, in the above formulation, each center has its own tuning parameter $h_c$. We use $h_c = \tilde{\sigma}_c n_c^{-1/3}$, where $\tilde{\sigma}_c$ is the sample standard error of $\tilde{\btheta}^{\top} \textbf{X}^c$. This mimics the approach used by others in similar work (e.g.,~\cite{lin2011}) and seems to work well. Since these parameters are estimated based on the starting value $\tilde{\btheta}$, the presence of $m$ tuning parameters does not present a large computational burden. Bootstrapping could be used to obtain confidence intervals for the estimated aAUC of the fitted combination $\hat{\btheta}$.

\subsection{Penalization}

In practice, it is unlikely that a given combination will have the same AUC in each center. This could be due to, for example, differences in the populations of patients at different centers. In this situation, it may be desirable to construct a biomarker combination that has relatively similar performance across centers. In particular, it may be worth sacrificing a small amount of the overall performance (in terms of the aAUC) for less variability in the center-specific AUCs. For instance, one combination may have an aAUC of 0.7 with center-specific AUCs ranging between 0.55 and 0.85; that is, in some centers the performance is very good, while in others it is quite poor. Another combination may have an aAUC of 0.68 with center-specific AUCs ranging between 0.64 and 0.72. This latter combination has slightly worse overall performance, but its discriminatory ability across centers is much less variable. In the typical multicenter setting, where only a fraction of the centers are sampled, this reduced variability may be desirable as it could lead to a fitted combination that has an AUC in new centers within a narrower range. 

To accomplish this, we propose
$\hat{\btheta}_{\lambda} = \argmaxA_{\btheta \in \boldsymbol{\Theta}} \left\lbrace aR_n(\btheta) - \lambda \sum_{c=1}^m \hat{w}_c \left(R^c_{n_c}(\btheta) - aR_n(\btheta) \right)^2  \right\rbrace,$
where $\lambda$ is a fixed penalty parameter, $\lambda \geq 0$. Since $\left\lbrace aR_n(\btheta) - \lambda \sum_{c=1}^m \hat{w}_c \left(R^c_{n_c}(\btheta) - aR_n(\btheta) \right)^2 \right\rbrace$ is the difference of two smooth functions, it can be maximized using gradient-based methods. The goal of this penalized method is to construct a combination such that its performance in a new center is similar to what has been observed in previous centers. The notion of ``similar'' depends upon the degree of underlying variability across the population of centers as well as the centers that have been sampled and are used to obtain $\hat{\btheta}_{\lambda}$. 

\subsection{Asymptotic results} \label{asymprslts}

In the theorem below, we demonstrate good operating characteristics for the combination $\hat{\btheta}_{\lambda}$ in large samples. By setting $\lambda = 0$, we obtain asymptotic results for the maximization of $aR_n(\btheta)$ without penalization. Define $\tilde{Q}_n(\btheta; \lambda) = aR_n(\btheta) - \lambda \sum_{c=1}^m \hat{w}_c\left(R_{n_c}^c(\btheta) - aR_n(\btheta)\right)^2$
and
$Q(\btheta; \lambda) = aAUC(\btheta) - \lambda \sum_{c=1}^M w_c \left(AUC_c(\btheta) - aAUC(\btheta)\right)^2.$
We first present several conditions necessary for the theorem:
\begin{enumerate}
\item[(A1)] The $m$ centers are randomly sampled from the population of $M$ centers and $n_c$ observations are randomly sampled from center $c$, $c=1,...,m$.
\item[(A2)] $\sum_{c=1}^m\left| E(\hat{w}_c) - w_c \right| \rightarrow 0$ as $n_c \rightarrow \infty$, $c = 1,...,m$, and $m \rightarrow M$ such that $\sqrt{n_c}/m \rightarrow \infty$.
\item[(A3)] The centers are independent and within each center, the observations $O^c_i = (D^c_i, \textbf{X}^c_i)$, $i=1,...,n_c$, are independent and identically distributed $(p+1)$-dimensional random vectors such that there exists at least one component of $\textbf{X}^c$, $X^{c}_k$ for some $k \in \lbrace 1, ..., p \rbrace$, with distribution that has everywhere positive Lebesgue density, conditional on the other $\textbf{X}^c$ components.
\item[(A4)] The support of $\textbf{X}^c$, $c = 1,...,M$, is not contained in any proper linear subspace of $\mathbb{R}^p$.
\item[(A5)] For fixed $\lambda \geq 0$, both the maximum of $\tilde{Q}_n(\btheta; \lambda)$ and the maximum of $Q(\btheta; \lambda)$ over $B = \lbrace \btheta \in \mathbb{R}^{p} : ||\btheta||=1, |\theta_k| > 0 \rbrace$ are attained.
\end{enumerate} 


\begin{theorem} 
Fix $\lambda \geq 0$ and suppose conditions (A1)--(A5) hold. Then \newline
$\sup_{\btheta \in B} Q(\btheta; \lambda) = Q(\hat{\btheta}_{\lambda}; \lambda) + o_p(1)$ as $n_c \rightarrow \infty$, $c = 1,...,m$, and $m \rightarrow M$ such that $\sqrt{n_c}/m \rightarrow \infty$. 
\end{theorem}

The proof of the theorem is given in Appendix A. We have previously demonstrated that, under certain conditions, $\hat{AUC}_c(\btheta)$ converges uniformly in probability to $AUC_c(\btheta)$ and $\hat{aAUC}(\btheta)$ converges uniformly in probability to $aAUC(\btheta)$, and we use these results in the proof of the theorem~\cite*{meisner2017}. Briefly, the proof first demonstrates uniform convergence in probability of the difference between $Q(\btheta; \lambda)$ and the empirical analogue of $\tilde{Q}_n(\btheta; \lambda)$ (that is, with $\hat{AUC}_c$ in place of $R^c_{n_c}$) to zero. The proof then uses previous results for $R_n$ to demonstrate uniform convergence in probability of the difference between $\tilde{Q}_n(\btheta; \lambda)$ and the empirical analogue of $\tilde{Q}_n(\btheta; \lambda)$ to zero. Combining these results gives the desired conclusion. 

This theorem does not consider the convergence of $\hat{\btheta}_{\lambda}$ to a particular quantity; instead, the theorem relates to the operating characteristics of the combination $\hat{\btheta}_{\lambda}^\top \textbf{X}$. By focusing on the performance of the fitted combination, rather than the combination itself, we remove the need to make additional (restrictive) assumptions. More importantly, since our goal is to develop a combination with high discriminatory ability, demonstrating good operating characteristics of the fitted combination is paramount.  

\subsection{Choosing the penalization parameter $\lambda$} \label{locosec}

In other penalized estimation procedures, such as ridge regression or the lasso, the penalty parameter $\lambda$ is typically chosen via cross-validation, where the value of $\lambda$ that gives the best cross-validated performance is selected. The motivation for cross-validation is that apparent measures of performance for a given model (that is, estimates of performance based on the data used to fit the model) tend to be optimistic~\cite*{hastie2016}. Cross-validation is one way of addressing this problem.

For our penalized estimation method, we can extend the ideas behind cross-validation to the multicenter setting. As just described, the goal of cross-validation is typically to get an idea of the performance in new observations. In the case of data from multiple centers, we would like to get an idea of the performance in new \textit{centers.} To that end, we propose the following procedure, which we call ``leave one center out cross-validation'' (LOCOCV):

\begin{enumerate}
\item Choose a sequence of $\lambda$ values: $\lbrace \lambda_1, \lambda_2, ..., \lambda_{r} \rbrace$
\item For each value of $\lambda$ and $i=1,...,m$:
\begin{enumerate}
\item Fit the biomarker combination using the data from all but the $i^{th}$ center.
\item Estimate the AUC of the fitted combination from (a) in the $i^{th}$ center.
\end{enumerate}
\item Plot the $m$ center-specific AUCs from (2b), the corresponding aAUC, and the variability in the center-specific AUCs around the aAUC in (i) the cross-validation ``training'' centers and (ii) the cross-validation ``test'' centers as a function of $\lambda$. 
\item Choose an appropriate value of $\lambda$, and use this value to fit the biomarker combination using the data from all $m$ centers.
\end{enumerate}

In some situations, it may be acceptable to sacrifice a small amount of overall performance in return for a substantial decrease in the variability of the center-specific AUCs. In other situations, any decline in overall performance may be undesirable. Consequently, we recommend using the cross-validation plot described above to choose $\lambda$, rather than proposing an automated procedure, as the relative costs and benefits of using a larger or smaller value of $\lambda$ depend on the specific context. In other words, individual users must determine the appropriate degree of penalization. 

\section{Results}

\subsection{Direct maximization without penalization ($\lambda = 0$)} \label{directrslts}

We used simulations to investigate the performance of the proposed direct maximization method. Most of these simulations were based on the set-up used by~\cite{fong2016}.

In each simulation, we generated a population of centers and individuals, and obtained training data by sampling from this population. In particular, we first sampled $m$ centers from the population of $M$ centers. Then, within each of the $m$ sampled centers, we sampled $n_c$ observations from the $N_c$ observations available in each center (where $N_c$ and $n_c$ did not vary across centers). These observations formed the training data, in which the combinations were constructed. The fitted combinations were then evaluated in independent test data, which consisted of the $N_c$ observations in each of the $M-m$ centers not used in the training data. We considered the following settings: (i) $M = 50, N_c = 5,000, m = 6, n_c = 200$, (ii) $M = 500, N_c = 500, m = 50, n_c = 50$, and (iii) $M = 5000, N_c = 200, m = 500, n_c = 20$.

\cite{fong2016} noted that the presence of outliers may lead to diminished performance of logistic regression, while methods based on maximizing the AUC may be less affected since the AUC is a rank-based measure. Thus, we considered simulations with and without outliers in the data-generating model. We focused on the setting of two biomarkers ($\textbf{X} = (X_1, X_2)$) and considered $\left( \textbf{X} \mid C \right) = \lbrace(1 - \Delta)\times \textbf{Z}_0 \rbrace + \lbrace \Delta \times \textbf{Z}_1 \rbrace$ and $(D \mid \textbf{X},C) \sim \mbox{Bernoulli}\left[f\lbrace \theta_0^C + 4 X_1 - 3 X_2 - (X_1 - X_2)^3 \rbrace \right],$
where $\textbf{Z}_0$ and $\textbf{Z}_1$ were independent bivariate normal random variables with mean zero and respective covariance matrices 
\begin{align*}
0.2\times \left(\begin{array}{cc}
1 & 0.9 \\
0.9 & 1 \end{array} \right),\;\; 2 \times \left(\begin{array}{cc}
1 & 0 \\
0 & 1 \end{array} \right),
\end{align*}
$f(v) = (1 + e^{-v})^{-1}$, $\theta_0^C \sim \mbox{Uniform}(-1,1)$, and $\Delta$ was an independent Bernoulli random variable with success probability $\pi$, where $\pi = 0.05$ when outliers were simulated and $\pi = 0$ otherwise. Other simulations with more complex center effects were considered, including those where the distribution of the biomarkers and/or the biomarker combination varied by center; the results were largely similar to those from the scenario described above. 

Estimates from robust logistic regression were used as starting values for the proposed SaAUC method, and the SaAUC method was compared to robust logistic regression and standard unconditional logistic regression, both with fixed center-specific intercepts. We used the robust logistic regression method proposed by~\cite{bianco1996}. This method uses a deviance function that limits the influence of individual observations on the model fit, making it more robust to outliers than standard logistic regression. In addition, when $m = 50$ or $m = 500$, we used conditional logistic regression both to provide starting values for and to compare with the SaAUC method. All methods fit a linear combination. The simulations were repeated 1000 times. 

The results are presented in Table~\ref{table1}. Clearly, when outliers are present, the proposed method outperforms standard and robust logistic regression, both in terms of the center-adjusted AUC and the center-specific AUCs. There also appears to be less variability in performance across simulations when the SaAUC approach is used to construct combinations. As was observed by~\cite{fong2016} for the AUC, when outliers were not present, the three methods produced combinations with comparable performance. In general, we found the results were very similar when conditional logistic regression was used to provide starting values for and as a comparison with the SaAUC method for $m = 50$ and $m = 500$ (Table 1 of Appendix B). 

\setlength{\tabcolsep}{3pt}

\begin{table}[ht]
\begin{center}
\def\~{\hphantom{0}}
\caption{Mean (standard deviation) of the aAUC in test data and mean (standard deviation) of the minimum and maximum center-specific AUCs ($AUC_c$) across the centers in the test data based on combinations fitted by logistic regression (GLM), robust logistic regression (rGLM), and the proposed method without penalization ($\lambda = 0$; SaAUC). Robust logistic regression estimates were used as the starting values for the SaAUC method.}
\label{table1}
\hspace*{-20mm}
\begin{tabular}{c@{\hskip 0.3in}ccc@{\hskip 0.3in}ccc@{\hskip 0.3in}ccc} 
\\ \hline
Outliers & $aAUC(\hat{\btheta}_{GLM})$ & \multicolumn{2}{c}{\hspace{-0.2in}$AUC_c(\hat{\btheta}_{GLM})$} & $aAUC(\hat{\btheta}_{rGLM})$ & \multicolumn{2}{c}{\hspace{-0.2in}$AUC_c(\hat{\btheta}_{rGLM})$} & $aAUC(\hat{\btheta}_{SaAUC})$ & \multicolumn{2}{c}{$AUC_c(\hat{\btheta}_{SaAUC})$} \\ 
& & Min & Max & & Min & Max & & Min & Max \\ \hline \\ [-9pt]
\multicolumn{10}{c}{$m=6$} \\ [2pt]
Yes & 0.6244 & 0.6065 & 0.6424 & 0.6492 & 0.6315 & 0.6666 & 0.6856 & 0.6684 & 0.7025 \\ 
 & (0.012) & (0.013) & (0.013) & (0.030) & (0.031) & (0.030) & (0.007) & (0.008) & (0.007) \\[4pt]
No & 0.7032 & 0.6866 & 0.7197 & 0.7032 & 0.6866 & 0.7196 & 0.7030 & 0.6864 & 0.7195 \\
 & (0.002) & (0.004) & (0.004) & (0.002) & (0.004) & (0.004) & (0.002) & (0.004) & (0.004) \\ \hline \\ [-9pt]
\multicolumn{10}{c}{$m = 50$} \\ [2pt]
Yes & 0.6233 & 0.5444 & 0.6992 & 0.6473 & 0.5692 & 0.7215 & 0.6843 & 0.6082 & 0.7564 \\
 & (0.008) & (0.014) & (0.012) & (0.027) & (0.030) & (0.026) & (0.004) & (0.011) & (0.009) \\ [4pt]
No & 0.7036 & 0.6301 & 0.7731 & 0.7036 & 0.6301 & 0.7731 & 0.7035 & 0.6299 & 0.7730 \\
 & (0.001) & (0.009) & (0.008) & (0.001) & (0.009) & (0.008) & (0.001) & (0.010) & (0.008) \\ \hline \\ [-9pt]
\multicolumn{10}{c}{$m = 500$} \\ [2pt]
Yes & 0.6221 & 0.4683 & 0.7659 & 0.6333 & 0.4798 & 0.7756 & 0.6796 & 0.5287 & 0.8154 \\
 & (0.004) & (0.015) & (0.013) & (0.013) & (0.020) & (0.017) & (0.004) & (0.015) & (0.012) \\ [4pt]
No & 0.7038 & 0.5574 & 0.8330 & 0.7038 & 0.5574 & 0.8330 & 0.7037 & 0.5573 & 0.8329 \\
 & (0.001) & (0.014) & (0.010) & (0.001) & (0.014) & (0.010) & (0.001) & (0.014) & (0.010) \\ \hline
\end{tabular}
\end{center}
\end{table}

The proposed SaAUC method had excellent convergence rates (fewer than $0.03\%$ of simulations failed). Robust logistic regression failed to converge in up to $3\%$ of simulations for $m=50$ and up to $15\%$ for $m=500$; when this happened, standard unconditional logistic regression was used to obtain starting values. In addition, when simulating data with outliers, in some instances the true biomarker combination was so large that it returned a non-value for the outcome $D$ (in \texttt{R}, this occurs for $(1 + e^{-v})^{-1}$ when $v > 800$). These observations were removed from the simulated dataset, though this affected fewer than $0.01\%$ of observations. Finally, for $m=500$, some of the training data centers were concordant (that is, all cases or all controls) and were removed from the analysis. Up to 11\% of simulations had one or two concordant centers in the training data. 

Beyond the setting of outliers in the data, we considered a scenario where the true biomarker combination involved ten biomarkers, but only two were available for fitting. As above, we generated a population of centers and individuals and obtained training data by sampling from this population, using the remaining centers as test data in which the fitted combinations were evaluated. In particular, we used $M = 50, N_c = 5,000, m = 6,$ and $n_c = 200$. Here, $\textbf{X} = (X_1,...,X_{10})$ and $(\textbf{X}|C) \sim N(\boldsymbol{\mu}_C, I_{10})$ where $I_{10}$ is the $10 \times 10$ identity matrix and $\boldsymbol{\mu}_c = -1$ for $c = 1,...,10$, $\boldsymbol{\mu}_c = 1$ for $c = 11,...,40$, and $\boldsymbol{\mu}_c = 0$ for $c = 41,...,50$. Center-specific intercepts $\alpha_0^C$ were drawn from a $\mbox{Uniform}(0.2, 0.8)$ distribution and $(D|\textbf{X},C)$ was generated as a Bernoulli random variable with success probability $f(\alpha_0^C + X_1^2 - 2X_2 + X_3 - 3X_4 + X_5 - 4X_6 + X_7 - X_8 + X_9 - X_{10})$, where $f(v) = (1+e^{-v})^{-1}$. Although the true combination is a function of all ten biomarkers, only $X_1$ and $X_2$ were available for data analysis. We used estimates from robust logistic regression as starting values for the proposed SaAUC method, and compared the SaAUC method to robust logistic regression and standard unconditional logistic regression. All methods fit a linear combination and the simulations were repeated 1000 times. The mean (standard deviation) aAUC in the test data for logistic regression, robust logistic regression, and the proposed SaAUC method were 0.6330 (0.018), 0.6284 (0.020), and 0.6453 (0.016), respectively. Thus, the proposed method offers a small gain in the aAUC over the logistic regression approaches. 

\subsection{Direct maximization with penalization ($\lambda > 0$)} \label{penrsltssec}

We explored our proposed penalized estimation procedure via simulated datasets. In particular, we used individual datasets generated under a variety of models to explore how the method may perform in practice. As was done in the earlier simulations, we first generated a population of centers and individuals and obtained training data by sampling from this population. Specifically, we considered a population of $M = 50$ centers with $N_c = 5,000$ observations in each and sampled $n_c = 200$ observations from each of $m=6$ centers to form the training data, with the observations in the remaining $M - m = 44$ centers serving as test data. 

We considered nearly 400 individual datasets; different data-generating mechanisms were used and included variations on the link function relating $(\textbf{X}, C)$ to $P(D=1|\textbf{X},C)$, the distribution of the biomarkers across centers, and the degree of heterogeneity in the true biomarker combination across centers. We simulated four independent normally distributed biomarkers with equal variance and throughout, the true biomarker combination in each center was linear. The details of the data-generating models for the examples presented here are given in the captions of the corresponding figures. Estimates from robust logistic regression were used as starting values for the penalized estimation procedure. For each simulation, we applied the LOCOCV procedure described in Section~\ref{locosec}. We considered 50 values of $\lambda$ equally-spaced (on the log scale) between 0.1 and 200. This range of values is somewhat arbitrary. In other penalized estimation procedures, it is common to choose the maximum value of $\lambda$ to be the value that returns coefficient estimates of 0. The analogous requirement in the current setting would be the value of $\lambda$ that gives center-specific AUCs of 0.5 in all centers. This is only expected to occur when all of the biomarker coefficients are 0, which cannot happen due to the constraint $||\btheta||=1$ in the penalized estimation procedure. The key point is that the range of $\lambda$ values used here is meant to be illustrative, not prescriptive. 

We present a handful of examples here, and include several more in Figures 1--16 of the Appendix C. All of the plots we present have the same layout: the left plot gives the training data results, the middle plot gives the results of the LOCOCV procedure, and the right plot gives the test data results. In each plot, the horizontal axis shows $\mbox{log}_{10}\lambda$. Throughout, the solid lines represent the results of the penalized estimation procedure, the dashed lines represent the standard logistic regression results, and the dot-dashed lines represent the robust logistic regression results. In each plot, the left vertical axis displays the AUC, and corresponds to the gray lines (center-specific AUCs for the penalized estimation procedure) and the black lines (center-adjusted AUCs for the penalized estimation procedure, robust logistic regression (``rGLM''), and standard logistic regression (``GLM'')). The right vertical axis shows the variability in the center-specific performance on the standard deviation scale and corresponds to the red lines (variability relative to the adjusted AUC in the training centers) and blue lines (variability relative to the adjusted AUC in the test centers). 

In the test data, the centers are so large that the AUCs calculated in these centers are presumed to be equal to the population values. In the training data and cross-validation procedure, on the other hand, the AUCs are empirical estimates. Thus, in the test data, for a combination $\hat{\btheta}$ fitted in the training data, the variability relative to the training centers is
$\sum_{c=1}^{M-m} w_c (AUC_c(\hat{\btheta}) - \hat{aAUC}(\hat{\btheta}))^2$ and the variability relative to the test centers is $\sum_{c=1}^{M-m} w_c (AUC_c(\hat{\btheta}) - aAUC(\hat{\btheta}))^2,$ where the $w_c$'s are the weights for the centers in the test data, $\hat{aAUC}$ is the adjusted AUC for the centers in the training data, and $aAUC$ and $AUC_c$ are the adjusted AUC and the center-specific AUC, respectively, for the centers in the test data. 

Figures 1 and 2 present examples where the LOCOCV procedure does a particularly nice job of mimicking the patterns in the test data. We encountered some datasets where the penalized estimation procedure did not work as well. For instance, in a small number of datasets, the variability increased with increasing $\lambda$ in the test data, despite the patterns seen in the training data and the LOCOCV results; Figure 3 presents one such example. In this situation, a value of $\lambda$ may be chosen that produces a fitted combination with worse overall performance and more variability in center-specific performance than would be obtained without penalization. However, in this example, the drop in overall performance is not large and the variability is fairly small even when $\lambda$ is large. Our simulations indicate that when the centers in the training data are not representative of the population of centers, the results from the cross-validation procedure may not reflect the patterns in the test data; such discrepancies would be expected in general when a resampling procedure is applied to a non-representative sample. 

\begin{figure}[ht]
\hspace{-7mm}
\includegraphics[width=7in]{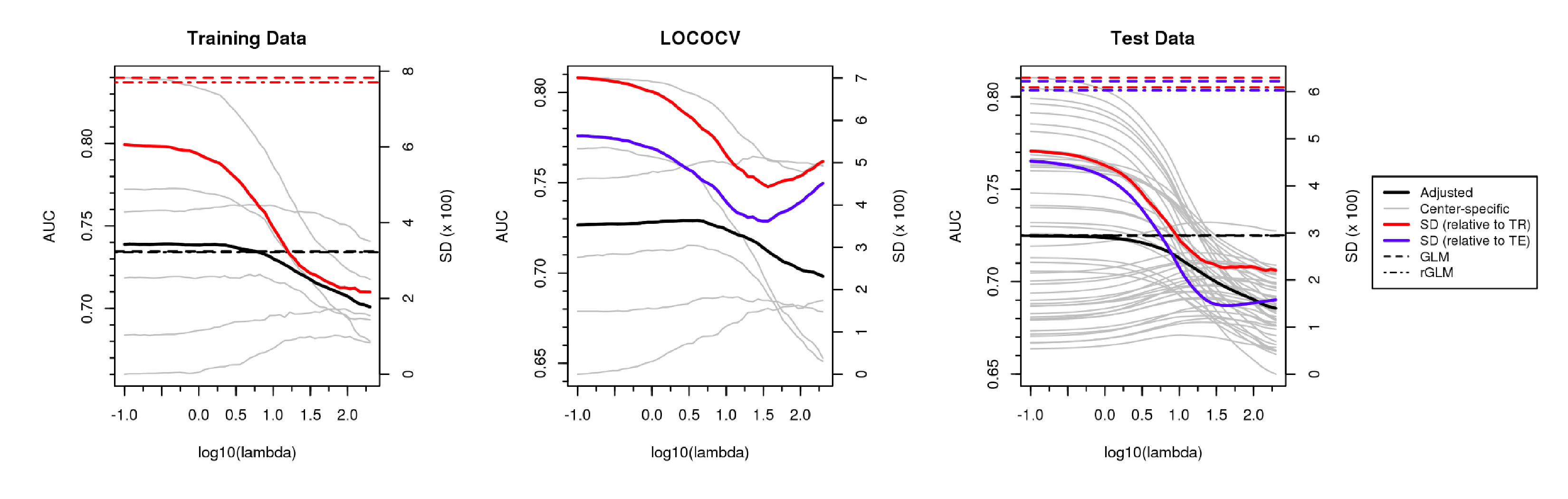}
\caption{Penalized estimation example 1. The four biomarkers $(X_1, X_2, X_3, X_4)$ were independently normally distributed with mean 0 and center-specific variances, i.e., $(X_i|C) \sim N(0, \sigma_{iC}^2), \;i=1,2,3,4$. The relationships between the biomarkers and $P(D=1)$ were allowed to vary by center and these variations (defined by $(\gamma_{1C}, \gamma_{2C}, \gamma_{3C}, \gamma_{4C})$) differed across the four biomarkers. The outcome $D$ was generated as a Bernoulli random variable with success probability $g(\alpha_0^C + \gamma_{1C}X_1 - \gamma_{2C}X_2 + \gamma_{3C}X_3 - \gamma_{4C}X_4$), where $\alpha_0^C$ is a center-specific intercept and $g(v) = (1+e^{-v/3})^{-1}$ for $v < 0$ and $(1+e^{-3v})^{-1}$ otherwise. In this example, the penalized estimation procedure produces combinations with reduced variability across centers with minimal loss in overall performance (for modest $\lambda$ values). Importantly, the LOCOCV results mimic what is seen in the test data.}
\label{pen1}
\end{figure}

\begin{figure}[ht]
\hspace{-7mm}
\includegraphics[width=7in]{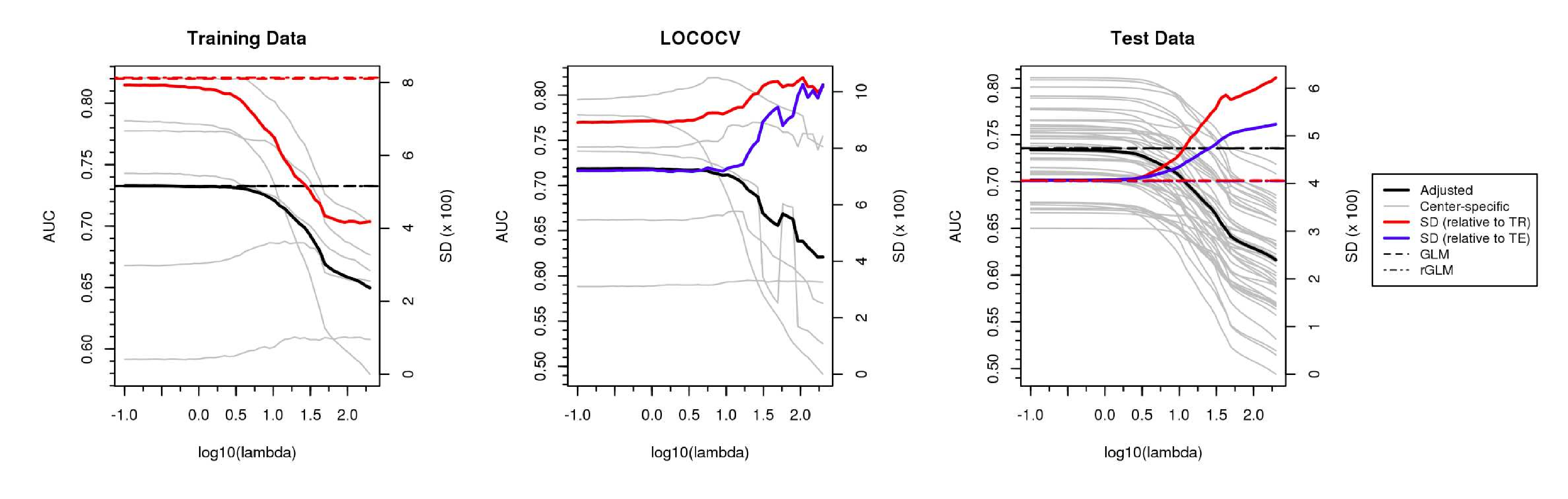}
\caption{Penalized estimation example 2. The four biomarkers $(X_1, X_2, X_3, X_4)$ were independently normally distributed with mean 0 and center-specific variances, i.e., $(X_i|C) \sim N(0, \sigma_{iC}^2), \;i=1,2,3,4$. The relationships between the biomarkers and $P(D=1)$ were allowed to vary by center and these variations (defined by $(\gamma_{1C}, \gamma_{2C}, \gamma_{3C}, \gamma_{4C})$) differed across the four biomarkers. The outcome $D$ was generated as a Bernoulli random variable with success probability $g(\alpha_0^C + \gamma_{1C}X_1 - \gamma_{2C}X_2 + \gamma_{3C}X_3 - \gamma_{4C}X_4$), where $\alpha_0^C$ is a center-specific intercept and $g(v) = (1+e^{-v/3})^{-1}$ for $v < 0$ and $(1+e^{-3v})^{-1}$ otherwise. This is an example where the LOCOCV results closely mimic the patterns seen in the test data, indicating the importance of performing cross-validation.}
\label{pen2}
\end{figure}

\begin{figure}[ht]
\hspace{-7mm}
\includegraphics[width=7in]{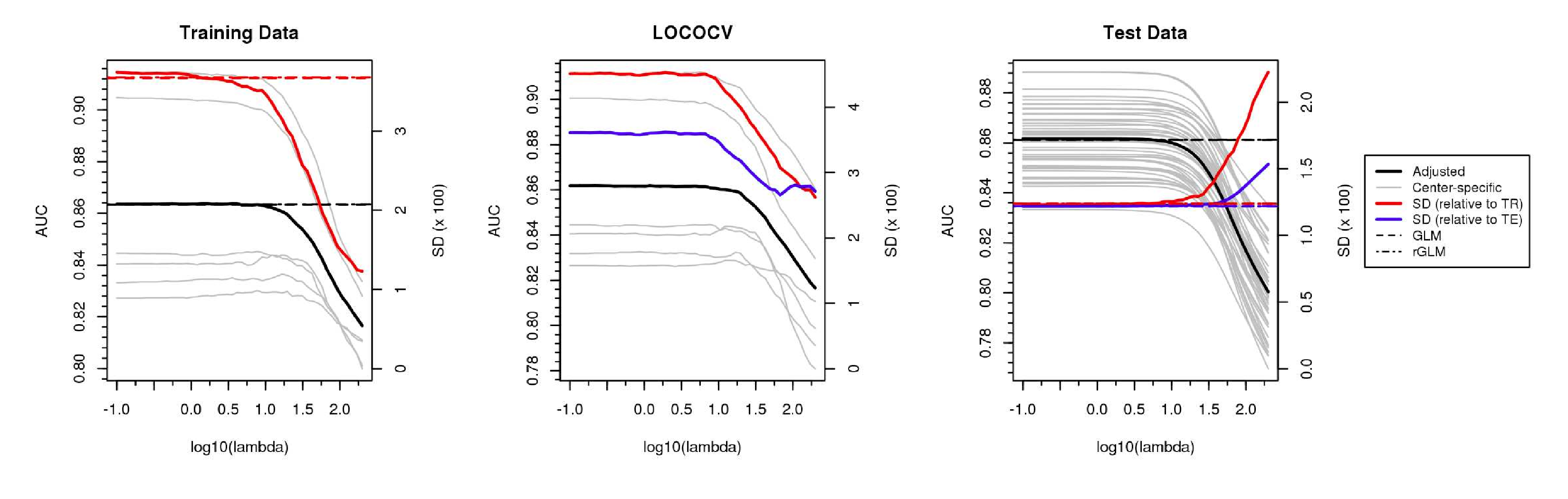}
\caption{Penalized estimation example 3. The four biomarkers $(X_1, X_2, X_3, X_4)$ were independently normally distributed with mean 0 and center-specific variances, i.e., $(X_i|C) \sim N(0, \sigma_{iC}^2), \;i=1,2,3,4$. The outcome $D$ was generated as a Bernoulli random variable with success probability $h(\alpha_0^C + X_1 - X_2 + X_3 - X_4$), where $\alpha_0^C$ is a center-specific intercept and $f(v) = (1+e^{-v})^{-1}$. This is an example where the penalization procedure does not work as well, since the variability in performance across centers increases with increasing $\lambda$, despite the patterns seen in the training data and the LOCOCV.}
\label{pen3}
\end{figure} 

Problems with convergence were encountered in fewer than 6\% of the examples we considered. Such issues generally arose with the more extreme scenarios we considered and primarily occurred during cross-validation. In practice, this may require modification of the range of $\lambda$ values considered. None of the results presented here had any convergence failures.  

\subsection{TRIBE-AKI study data}

To illustrate the methods we have developed, we used data from the TRIBE-AKI study and constructed combinations of three biomarkers measured no more than six hours after surgery: urine neutrophil gelatinase-associated lipocalin (NGAL), plasma heart-type fatty acid binding protein (h-FABP), and plasma troponin I (TNI). These data are used as illustration and not to report new findings of the TRIBE-AKI study. As described above, the TRIBE-AKI study involved individuals at six medical centers who were undergoing cardiac surgery and did not have evidence of AKI prior to surgery. AKI is a negative side effect of cardiac surgery, often caused by interruptions in blood flow to the kidneys during the surgery. However, AKI is currently diagnosed by assessing changes in serum creatinine, which frequently do not occur until several days after surgery. The aim of the TRIBE-AKI study was to use biomarkers measured soon after surgery to provide an earlier diagnosis of AKI. 

We removed observations with missing values for any of the three biomarkers (leaving 962 observations), log-transformed the biomarker values, and scaled the biomarkers to have equal variance to improve convergence of the gradient-based algorithm used by our methods. The prevalence of AKI in each center was between 7.8\% and 22.9\%, and the sizes of the centers ranged from 53 to 483 patients.

We applied standard logistic regression (``GLM''), robust logistic regression (``rGLM''), and the proposed SaAUC method to the TRIBE-AKI study data. The fitted combinations (with normalized coefficients) for GLM, rGLM, and SaAUC were 
\begin{align*}
0.0720*\mbox{NGAL} + 0.9917*\mbox{h-FABP} - 0.1068*\mbox{TNI} \\
0.0720*\mbox{NGAL} + 0.9917*\mbox{h-FABP} - 0.1068*\mbox{TNI} \\
0.0107*\mbox{NGAL} + 0.9585*\mbox{h-FABP} - 0.2849*\mbox{TNI},
\end{align*}
respectively. The apparent estimated center-adjusted AUCs for these combinations were 0.6878, 0.6878 and 0.6918, respectively. After correcting for resubstitution bias, the center-adjusted AUC estimates were 0.6819, 0.6820 and 0.6825. Thus, in these data, there was little difference in the performance of the combinations (though there were clear differences in the fitted combinations themselves). Furthermore, there was more optimistic bias in the apparent aAUC estimate for the combination fitted by the SaAUC method, which might be expected in general since the SaAUC method optimizes a smooth approximation to the estimated aAUC. 

Finally, we applied the proposed penalized estimation method to the TRIBE-AKI study data (Figure 4). The results from the LOCOCV procedure support choosing $\lambda \approx 10^{1.5}$, which is expected to give a reduction in variability in center-specific performance of about 25--30\%, with essentially no loss in overall (center-adjusted) performance. In particular, the LOCOCV results indicate that when $\lambda = 0.1$ (the smallest value considered by LOCOCV), the center-specific AUC estimates ranged from 0.6042 to 0.7250, but when $\lambda = 10^{1.5}$, the center-specific AUC estimates were between 0.6270 and 0.6986. Using $\lambda = 10^{1.5}$ in the full TRIBE-AKI study dataset yielded the combination $-0.1067*\mbox{NGAL} + 0.9911*\mbox{h-FABP} + 0.0798*\mbox{TNI}.$

\begin{figure}[ht]
\hspace{-7mm}
\includegraphics[width=7in]{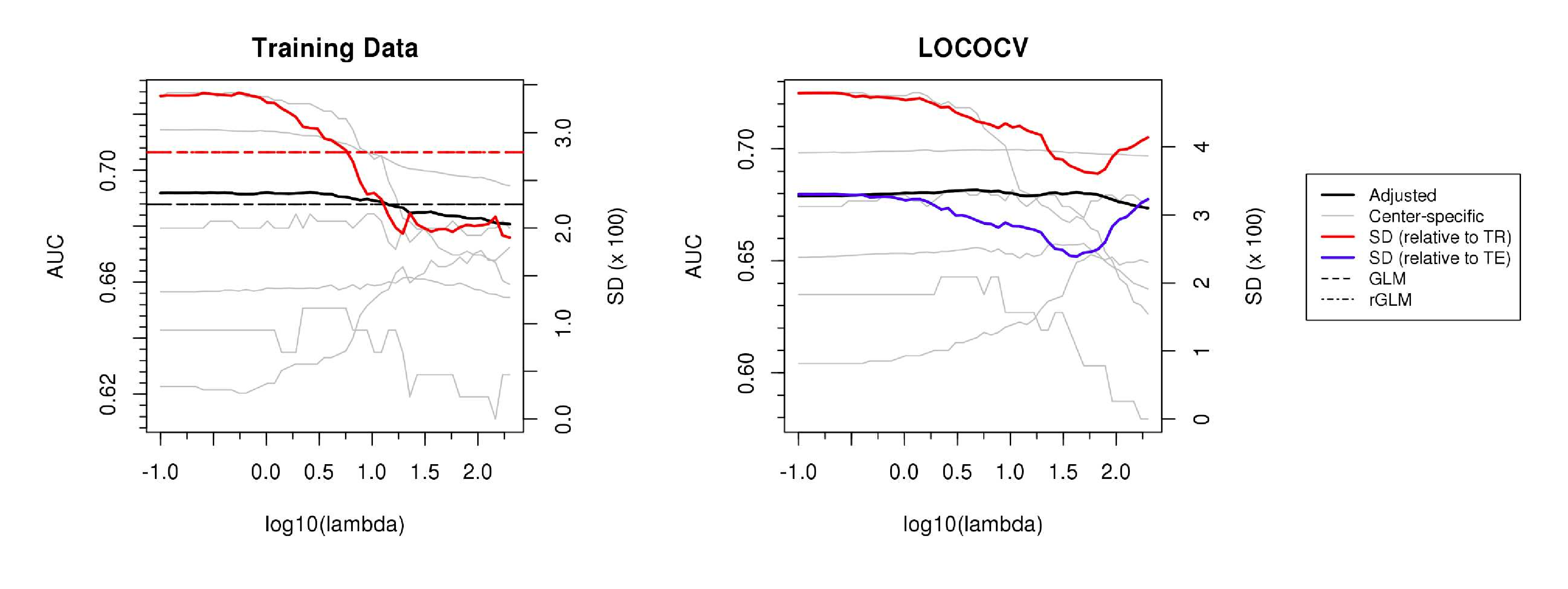}
\caption{Penalized estimation method applied to the TRIBE-AKI study data. The results from the LOCOCV procedure support choosing $\lambda \approx 10^{1.5}$, which is expected to give a reduction in variability in center-specific performance of about 25--30\%.}
\label{penTRIBE}
\end{figure} 

\section{Discussion}

We have proposed a method to construct biomarker combinations by maximizing a smooth approximation to the center-adjusted AUC. In addition, we have incorporated a penalty term that can be used to encourage similarity in performance across centers. Our method could be useful for other discrete nuisance covariates, such as batch, analysis laboratory, or measurement protocol. We used data on biomarkers measured after cardiac surgery to construct combinations for the diagnosis of acute kidney injury, demonstrating the feasibility of our methods. An \texttt{R} package including code to implement aAUC maximization without penalization, \texttt{maxadjAUC}, is publicly available via CRAN. In addition, \texttt{R} code to implement maximization with penalization is available on GitHub (\url{https://github.com/allisonmeisner/maxadjAUCpen}).

Multicenter studies include only a subset of centers, so predicted probabilities (risks) for individuals in new centers cannot be generated without relying on methods that require strong assumptions, e.g., random intercept logistic regression~\cite{meisner2017}. The methods proposed here do not provide predicted probabilities for individuals in either new centers or in the observed centers. Consequently, it is not possible to directly evaluate the calibration of a combination fitted by our method. Additionally, while we have provided a theoretical result related to the large-sample performance of combinations fitted by our method, we cannot offer guarantees on the performance of a fitted combination in a new center. We view this as a fundamental limitation of the data structure, rather than the methods. Moreover, when biomarkers exhibit ``center effects,'' further development of the biomarkers would be required prior to clinical application; after such development, predicted probabilities would change. A strength of the proposed methodology is that it does not rely on assumptions about the nature of ``center effects'' in constructing biomarker combinations; thus, it allows for the identification of promising combinations for further development while avoiding such assumptions.

In addition, in multicenter studies, different sampling schemes could be used (e.g., case-control or stratified case-control sampling). The estimated weights $\hat{w}_c$ would potentially be affected by different sampling procedures, and may not estimate $P(C=c|D=1)$. This would in turn affect the interpretation of the center-adjusted AUC, though it would still be a measure of the conditional performance. Our methods would then be optimizing this measure of conditional performance. The sampling scheme could also affect the validity of the asymptotic results we have provided. Furthermore, if a study involves matching, our methods would need to be modified to adjust the AUC for the matching in addition to center~\cite{janes2008}.

The adjusted AUC is a reasonable estimand even when the center-specific AUCs of a given combination are not the same across centers, though it is helpful to consider these center-specific AUCs, as they provide some insight into how the combination might be expected to perform in a new center. In addition, differences in performance across centers may be scientifically meaningful and merit further investigation. When assessing center-specific AUCs, it is important to also consider the sizes of the centers, as estimates of center-specific AUCs from small centers may be highly variable. One feature of our penalization approach is the use of the weights $\hat{w}_c$ in the penalty function, which reflect the proportion of cases in each center and so will tend to give less weight to small centers. Furthermore, the optimal combination (in terms of the center-specific AUC) may be different for each center. Importantly, however, our aim is not to identify the optimal combination in every center; instead, we are interested in constructing a single combination that performs well across centers. 

Since our smooth approximation function is not convex, further research is needed on the choice of starting values for the proposed method. It may also be possible to extend the method proposed by~\cite{fong2016}, which optimizes the convex ramp function approximation to the AUC, to the center-adjusted AUC. This may lead to further improvements in performance over logistic regression, as was seen in~\cite{fong2016} for the unadjusted AUC. In addition, when the centers are very small, the empirical center-specific AUC will be unreliable. Research is needed into the use of other (possibly parametric) methods to estimate the center-specific AUC by borrowing information across centers, which may be necessary when the centers are small. Furthermore, it may be possible to apply the ideas in this paper to survival data using the method proposed by~\cite{leborgne2017} for the covariate-adjusted AUC for censored data. Extending the proposed method to other center-adjusted measures of performance, such as the partial AUC or the true positive rate for a fixed false positive rate, is another avenue for future research. 

\section*{Acknowledgments}

This work was supported by the National Institutes of Health (F31 DK108356, R01 HL085757, and K24 DK090203). The opinions, results, and conclusions reported in this article are those of the authors and are independent of the funding sources.

The authors wish to acknowledge the TRIBE-AKI study investigators: Steven G. Coca (Department of Internal Medicine, Icahn School of Medicine at Mount Sinai, New York, New York), Amit X. Garg (Institute for Clinical Evaluative Sciences Western, London, Ontario, Canada; Division of Nephrology, Department of Medicine, and Department of Epidemiology and Biostatistics, University of Western Ontario, London, Ontario, Canada), Jay Koyner (Section of Nephrology, Department of Medicine, University of Chicago Pritzker School of Medicine, Chicago, Illinois), and Michael Shlipak (Kidney Health Research Collaborative, San Francisco Veterans Affairs Medical Center, University of California, San Francisco, San Francisco, California). Urine NGAL, plasma h-FABP, and plasma cardiac troponin I assays used in the TRIBE-AKI study were donated by Abbott Diagnostics, Randox Laboratories, and Beckman Coulter, respectively.

\section*{Data Availability Statement}

An \texttt{R} package containing code to implement aAUC maximization without penalization, \texttt{maxadjAUC}, is publicly available via CRAN. \texttt{R} code to implement maximization with penalization is available on GitHub (\url{https://github.com/allisonmeisner/maxadjAUCpen}).

\clearpage

\section*{Appendix A} 

\begin{proof}\textit{Proof of Theorem 1.}
First we will show $\sup_{\btheta \in B} \left| \tilde{Q}_n(\btheta; \lambda) - Q(\btheta; \lambda) \right| = o_p(1).$ Let \[ Q_n(\btheta; \lambda) = \hat{aAUC}(\btheta) - \lambda \sum_{c=1}^m \hat{w}_c\left(\hat{AUC}_c(\btheta) - \hat{aAUC}(\btheta)\right)^2.\] We can write
\begin{equation*}
\sup_{\btheta \in B} \left| \tilde{Q}_n(\btheta; \lambda) - Q(\btheta; \lambda) \right| \leq \sup_{\btheta \in B} \left| \tilde{Q}_n(\btheta; \lambda) - Q_n(\btheta; \lambda)\right| + \sup_{\btheta \in B} \left|Q_n(\btheta; \lambda) - Q(\btheta; \lambda) \right|.
\end{equation*}
Under conditions (A1)--(A4), we have shown (Lemma 1 and Theorem 1 in Meisner et al.~\cite{meisner2017}) that $ \sup_{\btheta \in B} \big| \hat{aAUC}(\btheta) - aAUC(\btheta) \big| = o_p(1)$ and \\
$\sup_{\btheta \in B} \left| \hat{AUC}_c(\btheta) - AUC_c(\btheta) \right| = o_p(1),\;\; c = 1,...,M.$ We can write
\begin{align*}
\sup_{\btheta \in B} & \left| Q_n(\btheta; \lambda) - Q(\btheta; \lambda) \right| \\
\leq & \sup_{\btheta \in B} \left| \hat{aAUC}(\btheta) - aAUC(\btheta) \right| \\
& + \lambda\sup_{\btheta \in B} \left| \sum_{c=1}^M w_c(AUC_c(\btheta) - aAUC(\btheta))^2 - \sum_{c=1}^m \hat{w}_c(\hat{AUC}_c(\btheta) - \hat{aAUC}(\btheta))^2 \right| \\
\leq & \sup_{\btheta \in B} \left| \hat{aAUC}(\btheta) - aAUC(\btheta) \right| \\
& + \lambda \sum_{c=m+1}^M \sup_{\btheta \in B} \left| w_c(AUC_c(\btheta) - aAUC(\btheta))^2 \right| \\
& + \lambda \sup_{\btheta \in B} \left| \sum_{c=1}^m \left\lbrace w_c(AUC_c(\btheta) - aAUC(\btheta))^2 - \hat{w}_c(\hat{AUC}_c(\btheta) - \hat{aAUC}(\btheta))^2 \right\rbrace \right|,
\end{align*}
where $\sum_{c=m+1}^M \sup_{\btheta \in B} \left| w_c(AUC_c(\btheta) - aAUC(\btheta))^2 \right| = o(1)$ as $m \rightarrow M$. Then by Theorem 1 in Meisner et al.~\cite{meisner2017},
\begin{align*}
\sup_{\btheta \in B} \left| Q_n(\btheta; \lambda) - Q(\btheta; \lambda) \right| \leq o_p(1) + o(1) + \lambda \sum_{c=1}^m  \sup_{\btheta \in B} \left| w_c Y^c_1(\btheta)^2 - \hat{w}_c(Y^c_2(\btheta) + Y^c_1(\btheta) + Y_3(\btheta))^2 \right|, 
\end{align*}
where $Y^c_1(\btheta) = AUC_c(\btheta) - aAUC(\btheta)$, $Y^c_2(\btheta) = \hat{AUC}_c(\btheta) - AUC_c(\btheta)$, and $Y_3(\btheta) = aAUC(\btheta) - \hat{aAUC}(\btheta)$; note that $|Y^c_1(\btheta)| \leq 1$, $|Y^c_2(\btheta)| \leq 1$ and $|Y_3(\btheta)| \leq 1$. Then
\begin{align*}
\sup_{\btheta \in B} & \left| Q_n(\btheta; \lambda) - Q(\btheta; \lambda) \right| \\
\leq \, & o_p(1) + o(1) + \lambda \sum_{c=1}^m  \sup_{\btheta \in B} \left| (w_c-\hat{w}_c) Y^c_1(\btheta)^2 \right. \\
& \left. - \hat{w}_c\left\lbrace Y^c_2(\btheta)^2 + Y_3(\btheta)^2 + 2 Y^c_1(\btheta)Y^c_2(\btheta) + 2 Y^c_1(\btheta)Y_3(\btheta) + 2 Y^c_2(\btheta)Y_3(\btheta) \right\rbrace \right| \\
\leq \, & o_p(1) + o(1) + \lambda \sum_{c=1}^m  \sup_{\btheta \in B} \left| (w_c-\hat{w}_c) Y^c_1(\btheta)^2 \right| + \lambda \sum_{c=1}^m  \sup_{\btheta \in B} \left| -\hat{w}_c Y^c_2(\btheta)^2  \right| \\
& + \lambda \sum_{c=1}^m  \sup_{\btheta \in B} \left| -\hat{w}_c Y_3(\btheta)^2  \right| + \lambda \sum_{c=1}^m  \sup_{\btheta \in B} \left| -2\hat{w}_c Y^c_1(\btheta)Y^c_2(\btheta)  \right| \\
& + \lambda \sum_{c=1}^m  \sup_{\btheta \in B} \left| -2\hat{w}_c Y^c_1(\btheta)Y_3(\btheta)  \right| + \lambda \sum_{c=1}^m  \sup_{\btheta \in B} \left| -2\hat{w}_c Y^c_2(\btheta)Y_3(\btheta)  \right|.
\end{align*}
We have (by Lemma 1 and Theorem 1 in Meisner et al.~\cite{meisner2017}) $\sup_{\btheta \in B} \left| Y^c_2(\btheta) \right| = o_p(1), c = 1,...,M$ and $\sup_{\btheta \in B} \left| Y_3(\btheta) \right| = o_p(1).$ This gives
\begin{align*}
\sup_{\btheta \in B} \left| (w_c-\hat{w}_c) \left[Y^c_1(\btheta)\right]^2 \right| & = \left| w_c-\hat{w}_c \right| \sup_{\btheta \in B} \left[ Y^c_1(\btheta) \right]^2  \leq \left| w_c-\hat{w}_c \right| \\ 
\sup_{\btheta \in B} \left| -\hat{w}_c \left[Y^c_2(\btheta)\right]^2 \right| & = \hat{w}_c \sup_{\btheta \in B} \left[Y^c_2(\btheta)\right]^2 \leq \hat{w}_c \sup_{\btheta \in B} \left| Y^c_2(\btheta) \right| = \hat{w}_c o_p(1) \\ 
\sup_{\btheta \in B} \left| -\hat{w}_c \left[ Y_3(\btheta) \right]^2 \right| & = \hat{w}_c \sup_{\btheta \in B} \left[ Y_3(\btheta) \right]^2 \leq \hat{w}_c \sup_{\btheta \in B} \left| Y_3(\btheta) \right| = \hat{w}_c o_p(1)  \\ 
\sup_{\btheta \in B} \left| -2\hat{w}_c Y^c_1(\btheta)Y^c_2(\btheta) \right| & = \hat{w}_c o_p(1) \\ 
\sup_{\btheta \in B} \left| -2\hat{w}_c Y^c_1(\btheta)Y_3(\btheta) \right| & = \hat{w}_c o_p(1) \\
\sup_{\btheta \in B} \left| -2\hat{w}_c Y^c_2(\btheta)Y_3(\btheta) \right| & = \hat{w}_c o_p(1).
\end{align*}
Since $\sum_{c=1}^m \hat{w}_c = 1$ for every $m$, 
\begin{equation*}
\sup_{\btheta \in B} \left| Q_n(\btheta; \lambda) - Q(\btheta; \lambda) \right| \leq o_p(1) + o(1) + \lambda \sum_{c=1}^m  \left| w_c-\hat{w}_c \right|.
\end{equation*}
Furthermore, we have previously shown (Theorem 1 in Meisner et al.~\cite{meisner2017}) that \\
$\sum_{c=1}^m  \left| w_c-\hat{w}_c \right| = o_p(1)$ as $n_c \rightarrow \infty$, $c = 1,...,m$, and $m \rightarrow M$ such that $\sqrt{n}_c/m \rightarrow \infty$. Thus, $\sup_{\btheta \in B} \left| Q_n(\btheta; \lambda) - Q(\btheta; \lambda) \right| = o_p(1).$

Now consider $\sup_{\btheta \in B} \left| \tilde{Q}_n(\btheta; \lambda) - Q_n(\btheta; \lambda) \right|$. We will first show 
$\sup_{\btheta \in B} \left| aR_n(\btheta) - \hat{aAUC}(\btheta) \right|$ $= o_p(1).$ Ma and Huang~\cite{ma2007} demonstrated that $\sup_{\btheta \in B} \left| R^c_{n_c}(\btheta) - \hat{AUC}_c(\btheta) \right| = o_p(1)$ as $n_c \rightarrow \infty$. We can write
\begin{equation*}
\sup_{\btheta \in B} \left| aR_n(\btheta) - \hat{aAUC}(\btheta) \right| \leq \sum_{c=1}^m \hat{w}_c \sup_{\btheta \in B} \left| R^c_{n_c}(\btheta) - \hat{AUC}_c(\btheta) \right| \leq \sum_{c=1}^m \hat{w}_c o_p(1) = o_p(1)
\end{equation*}
since $\sum_{c=1}^m \hat{w}_c = 1$ for every $m$.

Now consider $\sup_{\btheta \in B} \left| \tilde{Q}_n(\btheta; \lambda) - Q_n(\btheta; \lambda) \right|$. We can write
\begin{align*}
\sup_{\btheta \in B} &\left| \tilde{Q}_n(\btheta; \lambda) - Q_n(\btheta; \lambda) \right| \\
\leq \, & \sup_{\btheta \in B} \left| aR_n(\btheta) - \hat{aAUC}(\btheta) \right| \\
& + \lambda \sup_{\btheta \in B} \left| \sum_{c=1}^m \left\lbrace \hat{w}_c(\hat{AUC}_c(\btheta) - \hat{aAUC}(\btheta))^2 - \hat{w}_c(R^c_{n_c}(\btheta) - aR_n(\btheta))^2 \right\rbrace \right| \\
\leq \, & o_p(1) + \lambda \sum_{c=1}^m  \sup_{\btheta \in B} \left| \hat{w}_c Z^c_1(\btheta)^2 - \hat{w}_c(Z^c_2(\btheta) + Z^c_1(\btheta) + Z_3(\btheta))^2 \right|,
\end{align*}
where $Z^c_1(\btheta) = \hat{AUC}_c(\btheta) - \hat{aAUC}(\btheta)$, $Z^c_2(\btheta) = R^c_{n_c}(\btheta) - \hat{AUC}_c(\btheta)$, and $Z_3(\btheta) = \hat{aAUC}(\btheta) - aR_n(\btheta);$ note that $|Z^c_1(\btheta)| \leq 1$, $|Z^c_2(\btheta)| \leq 1$ and $|Z_3(\btheta)| \leq 1$. This gives
\begin{align*}
\sup_{\btheta \in B} \left| \tilde{Q}_n(\btheta; \lambda) - Q_n(\btheta; \lambda) \right| \leq \, & o_p(1) + \lambda \sum_{c=1}^m  \sup_{\btheta \in B} \left| -\hat{w}_c Z^c_2(\btheta)^2  \right| \\
& + \lambda \sum_{c=1}^m  \sup_{\btheta \in B} \left| -\hat{w}_c Z_3(\btheta)^2  \right| + \lambda \sum_{c=1}^m  \sup_{\btheta \in B} \left| -2\hat{w}_c Z^c_1(\btheta)Z^c_2(\btheta)  \right| \\
& + \lambda \sum_{c=1}^m  \sup_{\btheta \in B} \left| -2\hat{w}_c Z^c_1(\btheta)Z_3(\btheta)  \right| + \lambda \sum_{c=1}^m  \sup_{\btheta \in B} \left| -2\hat{w}_c Z^c_2(\btheta)Z_3(\btheta)  \right|.
\end{align*}
We have that $\sup_{\btheta \in B} \left| Z^c_2(\btheta) \right| = o_p(1), c = 1,...,M$ and $\sup_{\btheta \in B} \left| Z_3(\btheta) \right| = o_p(1).$
This gives
\begin{align*}
\sup_{\btheta \in B} \left| -\hat{w}_c \left[ Z^c_2(\btheta)\right]^2 \right| & = \hat{w}_c \sup_{\btheta \in B} \left[ Z^c_2(\btheta)\right]^2 \leq \hat{w}_c o_p(1) \\ 
\sup_{\btheta \in B} \left| -\hat{w}_c \left[ Z_3(\btheta)\right]^2 \right| & = \hat{w}_c \sup_{\btheta \in B} \left[Z_3(\btheta)\right]^2 \leq \hat{w}_c o_p(1)  \\ 
\sup_{\btheta \in B} \left| -2\hat{w}_c Z^c_1(\btheta)Z^c_2(\btheta) \right| & = \hat{w}_c o_p(1) \\ 
\sup_{\btheta \in B} \left| -2\hat{w}_c Z^c_1(\btheta)Z_3(\btheta) \right| & = \hat{w}_c o_p(1) \\
\sup_{\btheta \in B} \left| -2\hat{w}_c Z^c_2(\btheta)Z_3(\btheta) \right| & = \hat{w}_c o_p(1).
\end{align*}
Since $\sum_{c=1}^m \hat{w}_c = 1$ for every $m$, we have $\sup_{\btheta \in B} \left| \tilde{Q}_n(\btheta; \lambda) - Q_n(\btheta; \lambda) \right| = o_p(1).$

Combining these results, we have $\sup_{\btheta \in B} \left| \tilde{Q}_n(\btheta; \lambda) - Q(\btheta; \lambda) \right| = o_p(1).$ Then
\begin{align*}
\left| Q(\hat{\btheta}_{\lambda}; \lambda) - \sup_{\btheta \in B} Q(\btheta; \lambda) \right| \leq \, & \left| \sup_{\btheta \in B} Q(\btheta; \lambda) - \sup_{\btheta \in B} \tilde{Q}_n(\btheta; \lambda) \right| + \left| \sup_{\btheta \in B} \tilde{Q}_n(\btheta; \lambda) - Q(\hat{\btheta}_{\lambda}; \lambda) \right| \\
\leq \, & \sup_{\btheta \in B} \left| Q(\btheta; \lambda) - \tilde{Q}_n(\btheta; \lambda) \right| + \left| \tilde{Q}_n(\hat{\btheta}_{\lambda}; \lambda) - Q(\hat{\btheta}_{\lambda}; \lambda) \right| \\
\leq \, & o_p(1) + \sup_{\btheta \in B} \left| \tilde{Q}_n(\btheta; \lambda) - Q(\btheta; \lambda) \right| = o_p(1),
\end{align*}
giving $\sup_{\btheta \in B} Q(\btheta; \lambda) = Q(\hat{\btheta}_{\lambda}; \lambda) + o_p(1)$ as $n_c \rightarrow \infty$, $c = 1,...,m$, and $m \rightarrow M$ such that $\sqrt{n}_c/m \rightarrow \infty$. 
\end{proof} 

\clearpage 

\section*{Appendix B}
 
\begin{table}[ht]
\begin{center}
\def~{\hphantom{0}}
\captionsetup{labelformat=empty}
\caption{\textbf{Table 1:} Mean (standard deviation) of the aAUC in test data and mean (standard deviation) of the minimum and maximum center-specific AUCs ($AUC_c$) across the centers in the test data based on combinations fitted by conditional logistic regression (GLM) and the proposed method without penalization ($\lambda = 0$; SaAUC). Conditional logistic regression estimates were used as the starting values for the SaAUC method.}{
\begin{tabular}{c@{\hskip 0.3in}ccc@{\hskip 0.3in}ccc}
\\ \hline
Outliers & $aAUC(\hat{\btheta}_{GLM})$ & \multicolumn{2}{c}{\hspace{-0.2in}$AUC_c(\hat{\btheta}_{GLM})$} & $aAUC(\hat{\btheta}_{SaAUC})$ & \multicolumn{2}{c}{\hspace{0.05in}$AUC_c(\hat{\btheta}_{SaAUC})$} \\ 
& & Min & Max & & Min & Max \\ \hline \\ [-9pt]
\multicolumn{7}{c}{$m = 50$} \\ [2pt]
Yes & 0.6233 & 0.5444 & 0.6992 & 0.6824 & 0.6062 & 0.7547 \\ 
 & (0.008) & (0.014) & (0.012) & (0.004) & (0.011) & (0.009) \\ [4pt]
No & 0.7036 & 0.6301 & 0.7731 & 0.7035 & 0.6299 & 0.7730 \\
 & (0.001) & (0.009) & (0.008) & (0.001) & (0.010) & (0.008) \\ \hline \\ [-9pt]
\multicolumn{7}{c}{$m = 500$} \\ [2pt]
Yes & 0.6221 & 0.4684 & 0.7659 & 0.6764 & 0.5253 & 0.8128 \\
 & (0.004) & (0.015) & (0.013) & (0.003) & (0.015) & (0.012) \\ [4pt]
No & 0.7038 & 0.5574 & 0.8330 & 0.7037 & 0.5573 & 0.8329 \\ 
 & (0.001) & (0.014) & (0.010) & (0.001) & (0.014) & (0.010) \\ \hline
\end{tabular}}
\end{center}
\end{table}


\clearpage

\section*{Appendix C} 

\begin{figure}[ht]
\hspace{-7mm}
\includegraphics[width=7in]{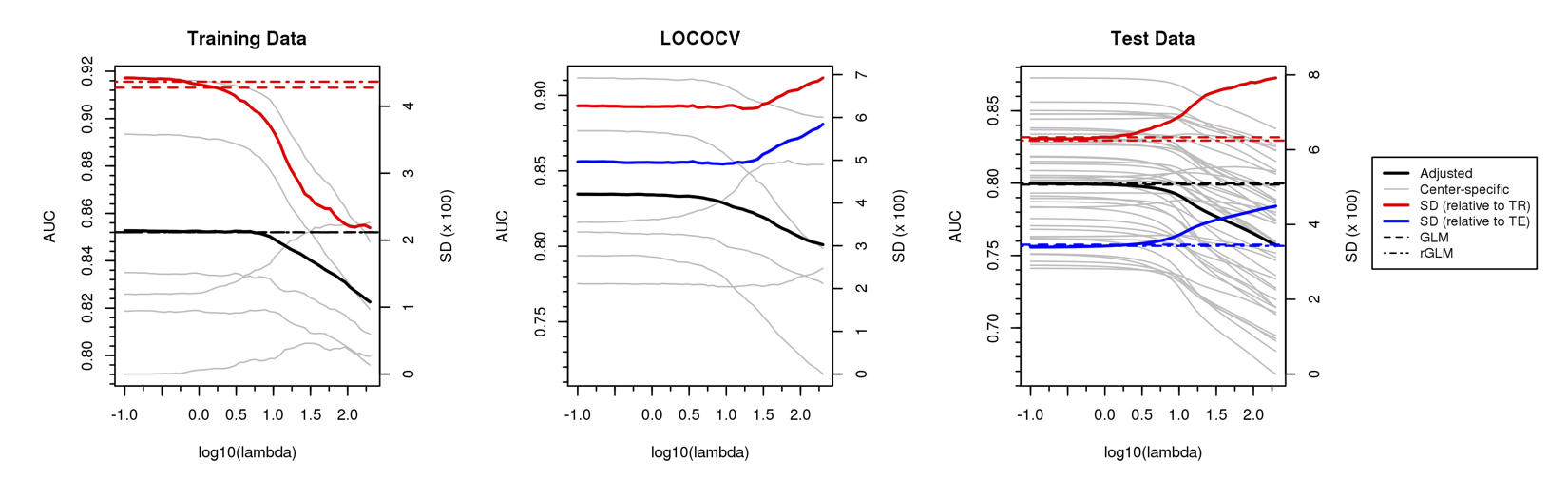}
\captionsetup{labelformat=empty}
\caption{\textbf{Figure 1:} Penalized estimation example 4. The four biomarkers $(X_1, X_2, X_3, X_4)$ were independently normally distributed with mean 0 and center-specific variances, i.e., $(X_i|C) \sim N(0, \sigma_{iC}^2), \;i=1,2,3,4$. The relationships between the biomarkers and $P(D=1)$ were allowed to vary by center and these variations (defined by $(\gamma_{1C}, \gamma_{2C}, \gamma_{3C}, \gamma_{4C})$) differed across the four biomarkers. The outcome $D$ was generated as a Bernoulli random variable with success probability $f(\alpha_0^C + \gamma_{1C}X_1 - \gamma_{2C}X_2 + \gamma_{3C}X_3 - \gamma_{4C}X_4$), where $\alpha_0^C$ is a center-specific intercept and $f(v) = (1+e^{-v})^{-1}$. This is an example where the LOCOCV procedure does very well in mimicking the patterns seen in the test data.}
\end{figure}

\begin{figure}[ht]
\hspace{-7mm}
\includegraphics[width=7in]{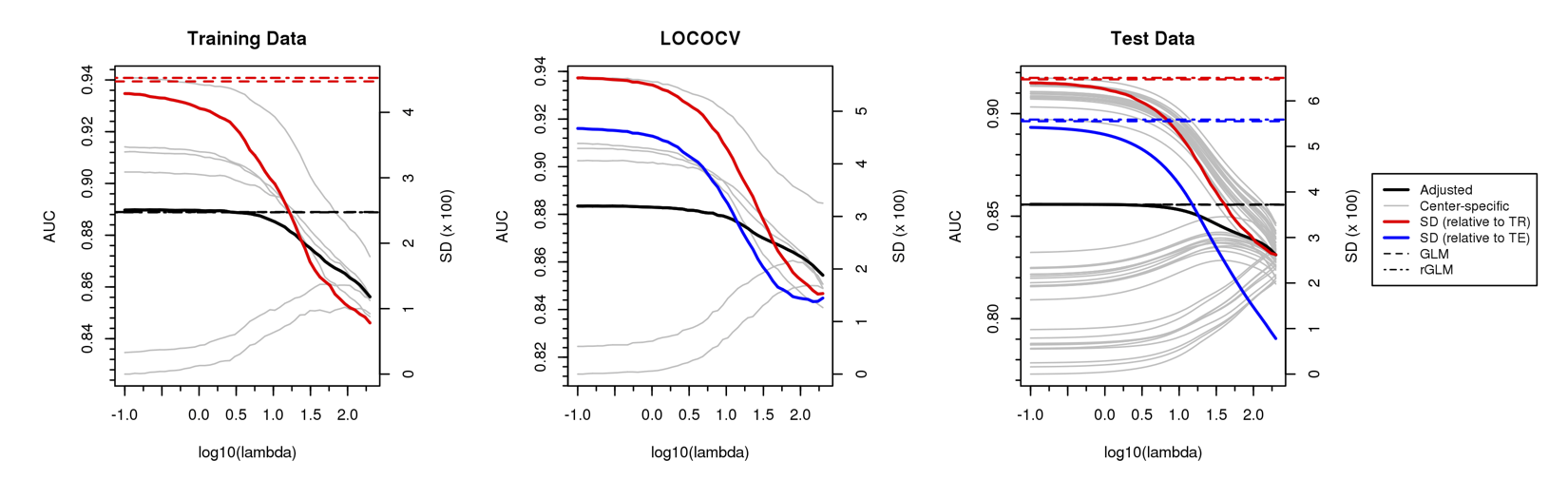}
\captionsetup{labelformat=empty}
\caption{\textbf{Figure 2:} Penalized estimation example 5. The four biomarkers $(X_1, X_2, X_3, X_4)$ were independently normally distributed with mean 0 and center-specific variances, i.e., $(X_i|C) \sim N(0, \sigma_{C}^2), \;i=1,2,3,4$. The relationships between the biomarkers and $P(D=1)$ were allowed to vary by center and these variations (defined by $(\gamma_{1C}, \gamma_{2C}, \gamma_{3C}, \gamma_{4C})$) differed across the four biomarkers. The outcome $D$ was generated as a Bernoulli random variable with success probability $f(\alpha_0^C + \gamma_{1C}X_1 - \gamma_{2C}X_2 + \gamma_{3C}X_3 - \gamma_{4C}X_4$), where $\alpha_0^C$ is a center-specific intercept and $f(v) = (1+e^{-v})^{-1}$. This illustrates a setting where there is a clear benefit to penalizing as there is a reduction in variability with little loss in overall performance.}
\end{figure}

\begin{figure}[ht]
\hspace{-7mm}
\includegraphics[width=7in]{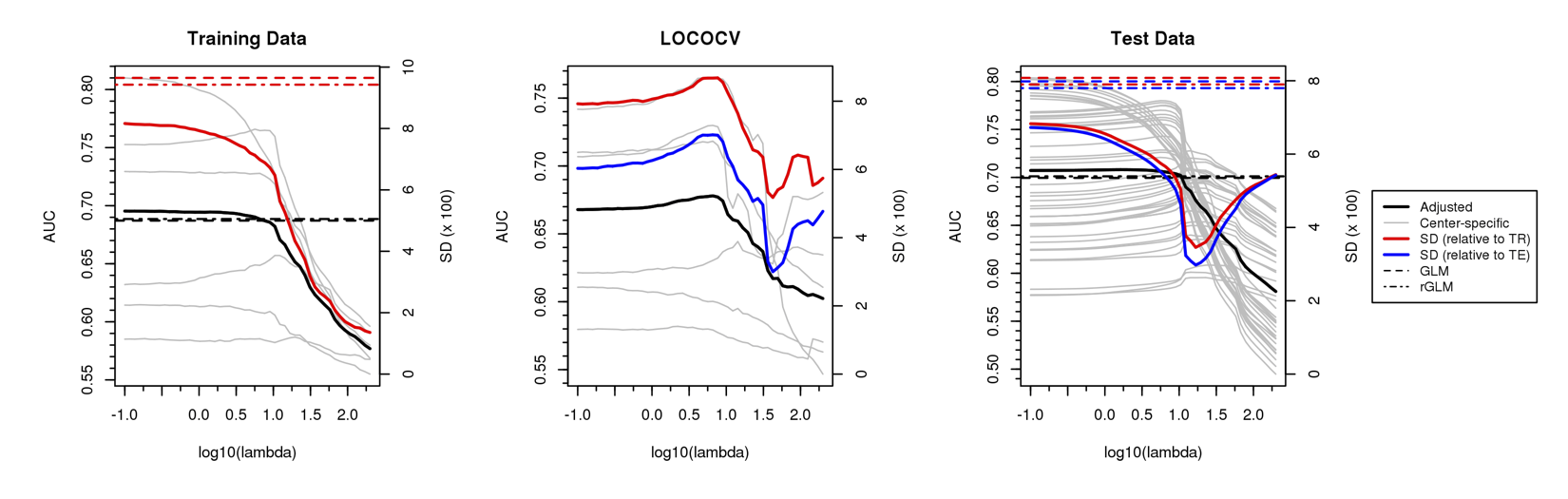}
\captionsetup{labelformat=empty}
\caption{\textbf{Figure 3:} Penalized estimation example 6. The four biomarkers $(X_1, X_2, X_3, X_4)$ were independently normally distributed with mean 0 and center-specific variances, i.e., $(X_i|C) \sim N(0, \sigma_{C}^2), \;i=1,2,3,4$. The relationships between the biomarkers and $P(D=1)$ were allowed to vary by center and these variations (defined by $(\gamma_{1C}, \gamma_{2C}, \gamma_{3C}, \gamma_{4C})$) differed across the four biomarkers. The outcome $D$ was generated as a Bernoulli random variable with success probability $g(\alpha_0^C + \gamma_{1C}X_1 - \gamma_{2C}X_2 + \gamma_{3C}X_3 - \gamma_{4C}X_4$), where $\alpha_0^C$ is a center-specific intercept and $g(v) = (1+e^{-v/3})^{-1}$ for $v < 0$ and $(1+e^{-3v})^{-1}$ otherwise. This is an example where the LOCOCV results are inconclusive in terms of which value of $\lambda$ should be chosen.}
\end{figure}

\begin{figure}[ht]
\hspace{-7mm}
\includegraphics[width=7in]{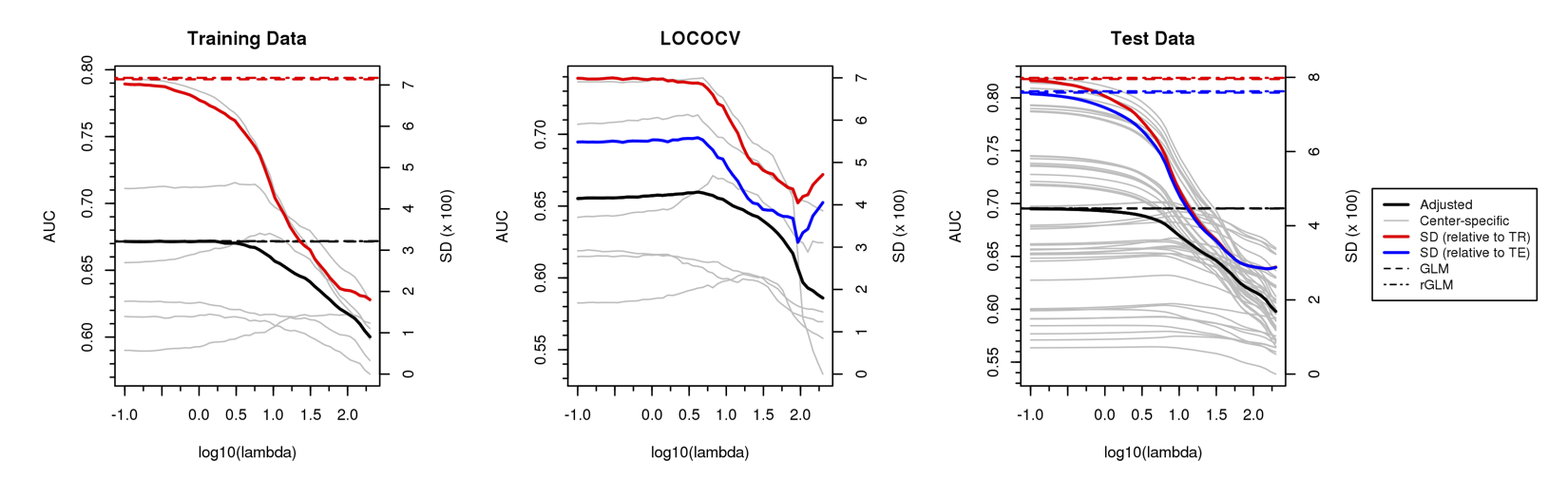}
\captionsetup{labelformat=empty}
\caption{\textbf{Figure 4:} Penalized estimation example 7. The four biomarkers $(X_1, X_2, X_3, X_4)$ were independently normally distributed with mean 0 and center-specific variances, i.e., $(X_i|C) \sim N(0, \sigma_{iC}^2), \;i=1,2,3,4$. The relationships between the biomarkers and $P(D=1)$ were allowed to vary by center and these variations (defined by $(\gamma_{1C}, \gamma_{2C}, \gamma_{3C}, \gamma_{4C})$) differed across the four biomarkers. The outcome $D$ was generated as a Bernoulli random variable with success probability $g(\alpha_0^C + \gamma_{1C}X_1 - \gamma_{2C}X_2 + \gamma_{3C}X_3 - \gamma_{4C}X_4$), where $\alpha_0^C$ is a center-specific intercept and $g(v) = (1+e^{-v/3})^{-1}$ for $v < 0$ and $(1+e^{-3v})^{-1}$ otherwise. This is an example where the penalization procedure does not work as well, since in the test data, the aAUC decreased more quickly with increasing $\lambda$ than was indicated by the LOCOCV procedure and the training data.}
\end{figure}

\begin{figure}[ht]
\hspace{-7mm}
\includegraphics[width=7in]{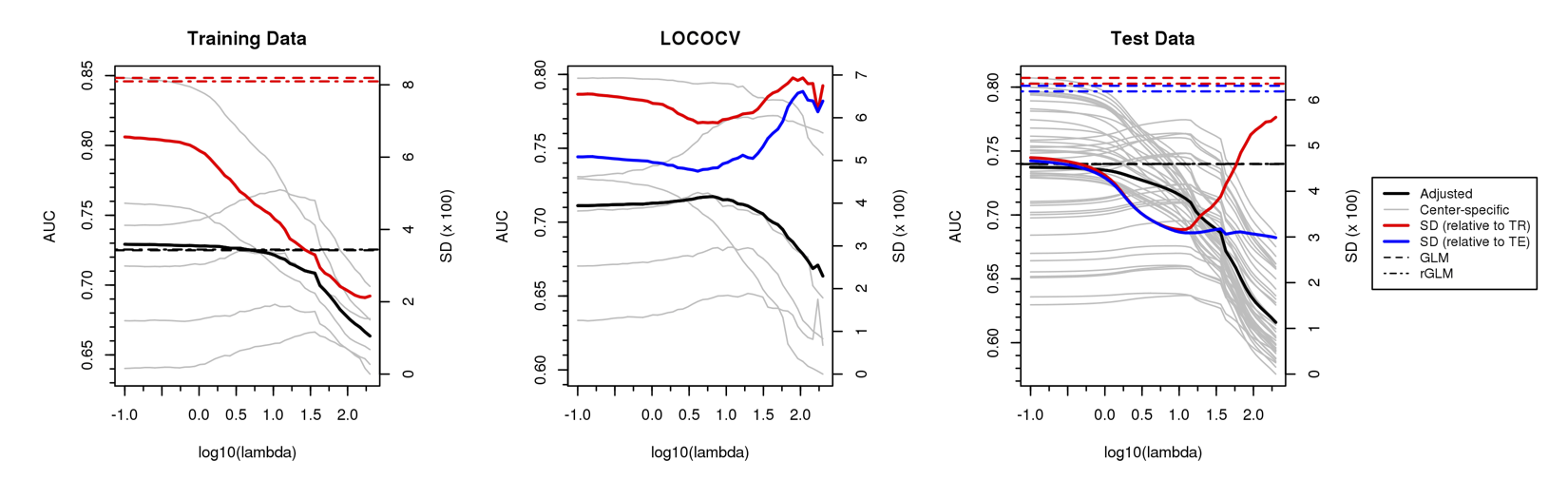}
\captionsetup{labelformat=empty}
\caption{\textbf{Figure 5:} Penalized estimation example 8. The four biomarkers $(X_1, X_2, X_3, X_4)$ were independently normally distributed with mean 0 and center-specific variances, i.e., $(X_i|C) \sim N(0, \sigma_{iC}^2), \;i=1,2,3,4$. The relationships between the biomarkers and $P(D=1)$ were allowed to vary by center and these variations (defined by $(\gamma_{1C}, \gamma_{2C}, \gamma_{3C}, \gamma_{4C})$) differed across the four biomarkers. The outcome $D$ was generated as a Bernoulli random variable with success probability $g(\alpha_0^C + \gamma_{1C}X_1 - \gamma_{2C}X_2 + \gamma_{3C}X_3 - \gamma_{4C}X_4$), where $\alpha_0^C$ is a center-specific intercept and $g(v) = (1+e^{-v/3})^{-1}$ for $v < 0$ and $(1+e^{-3v})^{-1}$ otherwise. In this example, the LOCOCV procedure does a nice job capturing the trends seen in the test data.}
\end{figure}

\begin{figure}[ht]
\hspace{-7mm}
\includegraphics[width=7in]{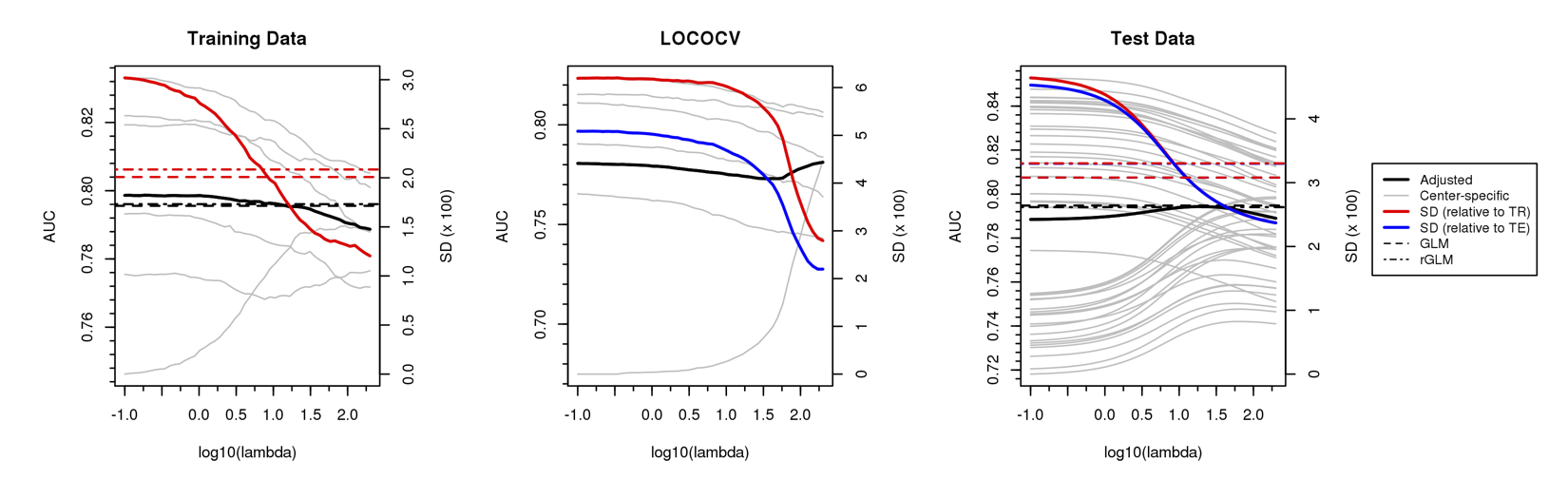}
\captionsetup{labelformat=empty}
\caption{\textbf{Figure 6:} Penalized estimation example 9. The four biomarkers $(X_1, X_2, X_3, X_4)$ were independently normally distributed with mean 0 and center-specific variances, i.e., $(X_i|C) \sim N(0, \sigma_{C}^2), \;i=1,2,3,4$. The relationships between the biomarkers and $P(D=1)$ were allowed to vary by center and these variations (defined by $(\gamma_{1C}, \gamma_{2C}, \gamma_{3C}, \gamma_{4C})$) differed across the four biomarkers. The outcome $D$ was generated as a Bernoulli random variable with success probability $g(\alpha_0^C + \gamma_{1C}X_1 - \gamma_{2C}X_2 + \gamma_{3C}X_3 - \gamma_{4C}X_4$), where $\alpha_0^C$ is a center-specific intercept and $g(v) = (1+e^{-v/3})^{-1}$ for $v < 0$ and $(1+e^{-3v})^{-1}$ otherwise. In this example, there is a clear benefit to penalization.}
\end{figure}

\begin{figure}[ht]
\hspace{-7mm}
\includegraphics[width=7in]{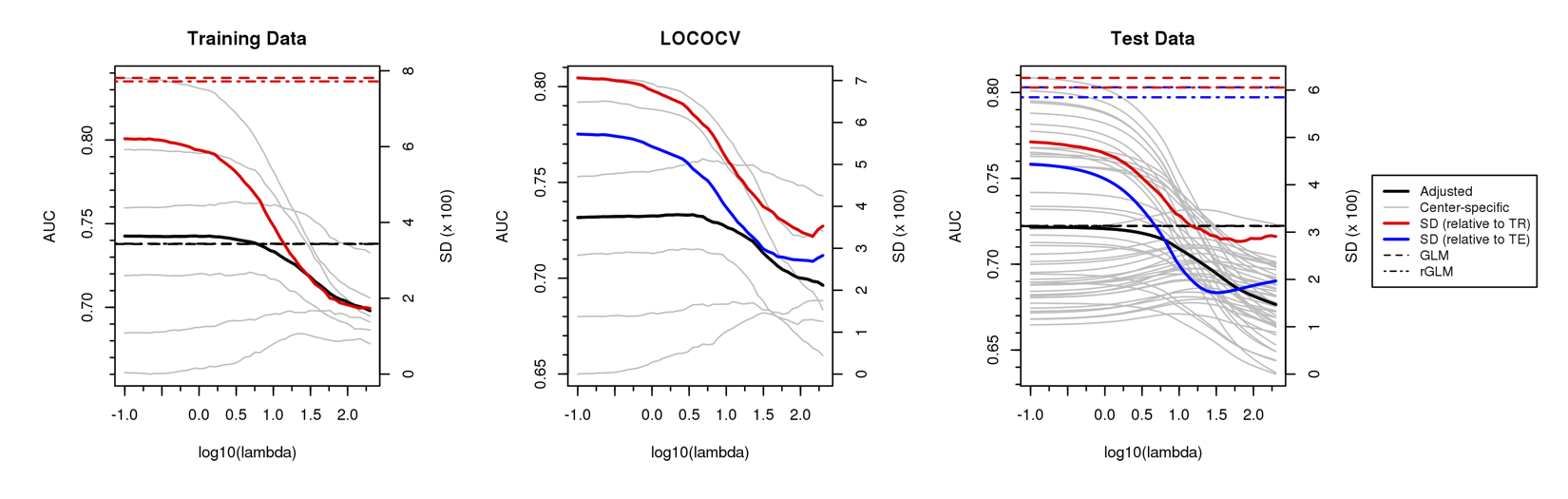}
\captionsetup{labelformat=empty}
\caption{\textbf{Figure 7:} Penalized estimation example 10. The four biomarkers $(X_1, X_2, X_3, X_4)$ were independently normally distributed with mean 0 and center-specific variances, i.e., $(X_i|C) \sim N(0, \sigma_{iC}^2), \;i=1,2,3,4$. The relationships between the biomarkers and $P(D=1)$ were allowed to vary by center and these variations (defined by $(\gamma_{1C}, \gamma_{2C}, \gamma_{3C}, \gamma_{4C})$) differed across the four biomarkers. The outcome $D$ was generated as a Bernoulli random variable with success probability $g(\alpha_0^C + \gamma_{1C}X_1 - \gamma_{2C}X_2 + \gamma_{3C}X_3 - \gamma_{4C}X_4$), where $\alpha_0^C$ is a center-specific intercept and $g(v) = (1+e^{-v/3})^{-1}$ for $v < 0$ and $(1+e^{-3v})^{-1}$ otherwise. In this example, there is a clear benefit to penalization for a range of $\lambda$ values.}
\end{figure}

\begin{figure}[ht]
\hspace{-7mm}
\includegraphics[width=7in]{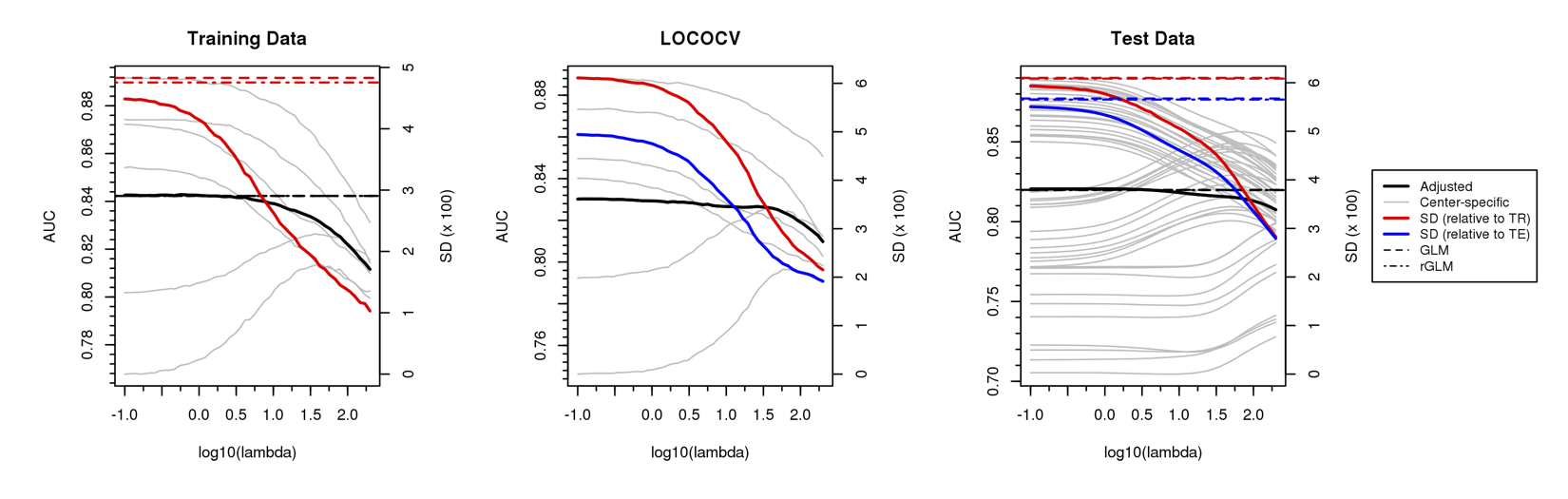}
\captionsetup{labelformat=empty}
\caption{\textbf{Figure 8:} Penalized estimation example 11. The four biomarkers $(X_1, X_2, X_3, X_4)$ were independently normally distributed with mean 0 and center-specific variances, i.e., $(X_i|C) \sim N(0, \sigma_{C}^2), \;i=1,2,3,4$. The relationships between the biomarkers and $P(D=1)$ were allowed to vary by center and these variations (defined by $(\gamma_{1C}, \gamma_{2C}, \gamma_{3C}, \gamma_{4C})$) differed across the four biomarkers. The outcome $D$ was generated as a Bernoulli random variable with success probability $g(\alpha_0^C + \gamma_{1C}X_1 - \gamma_{2C}X_2 + \gamma_{3C}X_3 - \gamma_{4C}X_4$), where $\alpha_0^C$ is a center-specific intercept and $g(v) = (1+e^{-v/3})^{-1}$ for $v < 0$ and $(1+e^{-3v})^{-1}$ otherwise. In this example, there is a clear benefit to penalization.}
\end{figure}

\begin{figure}[ht]
\hspace{-7mm}
\includegraphics[width=7in]{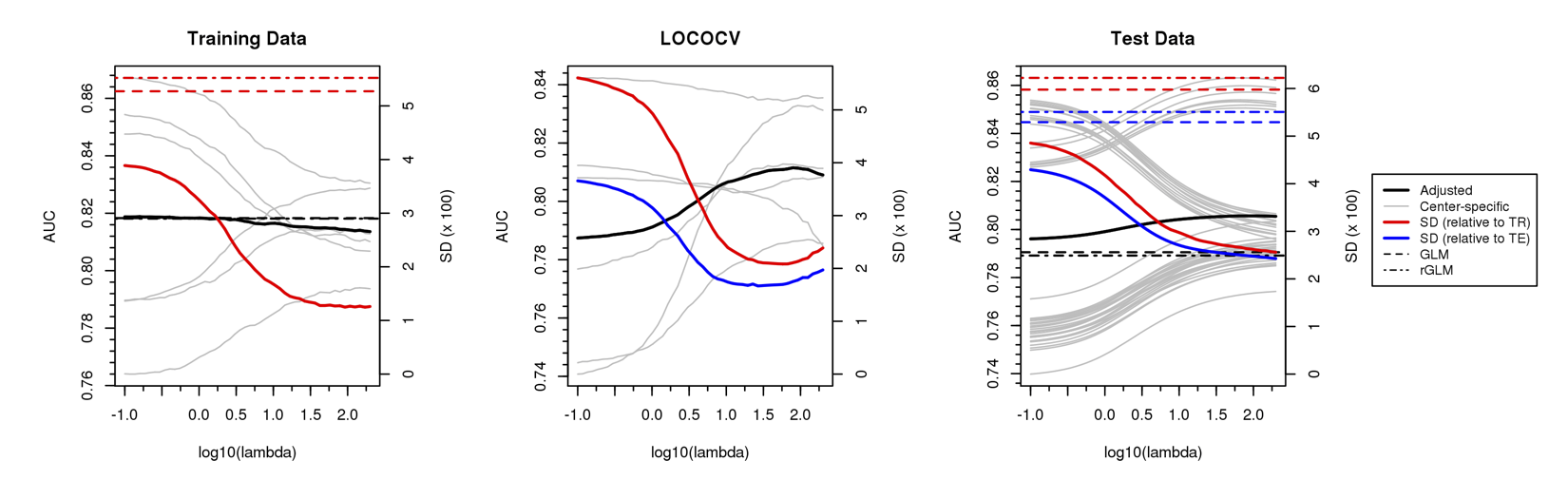}
\captionsetup{labelformat=empty}
\caption{\textbf{Figure 9:} Penalized estimation example 12. The four biomarkers $(X_1, X_2, X_3, X_4)$ were independently normally distributed with mean 0 and center-specific variances, i.e., $(X_i|C) \sim N(0, \sigma_{C}^2), \;i=1,2,3,4$. The relationships between the biomarkers and $P(D=1)$ were allowed to vary by center and these variations (defined by $(\gamma_{1C}, \gamma_{2C}, \gamma_{3C}, \gamma_{4C})$) differed across the four biomarkers. The outcome $D$ was generated as a Bernoulli random variable with success probability $f(\alpha_0^C + \gamma_{1C}X_1 - \gamma_{2C}X_2 + \gamma_{3C}X_3 - \gamma_{4C}X_4$), where $\alpha_0^C$ is a center-specific intercept and $f(v) = (1+e^{-v})^{-1}$. Here, there is a clear benefit to penalization, and the overall performance even increases slightly as $\lambda$ increases.}
\end{figure}

\begin{figure}[ht]
\hspace{-7mm}
\includegraphics[width=7in]{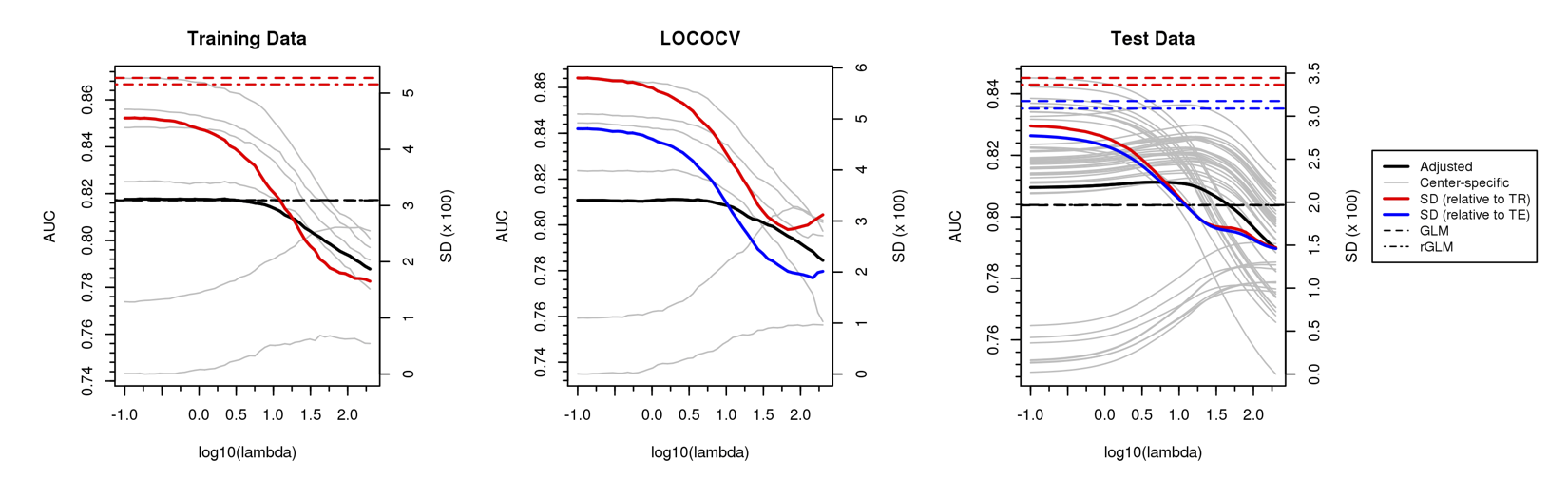}
\captionsetup{labelformat=empty}
\caption{\textbf{Figure 10:} Penalized estimation example 13. The four biomarkers $(X_1, X_2, X_3, X_4)$ were independently normally distributed with mean 0 and center-specific variances, i.e., $(X_i|C) \sim N(0, \sigma_{C}^2), \;i=1,2,3,4$. The relationships between the biomarkers and $P(D=1)$ were allowed to vary by center and these variations (defined by $(\gamma_{1C}, \gamma_{2C}, \gamma_{3C}, \gamma_{4C})$) differed across the four biomarkers. The outcome $D$ was generated as a Bernoulli random variable with success probability $f(\alpha_0^C + \gamma_{1C}X_1 - \gamma_{2C}X_2 + \gamma_{3C}X_3 - \gamma_{4C}X_4$), where $\alpha_0^C$ is a center-specific intercept and $f(v) = (1+e^{-v})^{-1}$. In this example, there is a definite benefit to penalization for a range of $\lambda$ values.}
\end{figure}

\begin{figure}[ht]
\hspace{-7mm}
\includegraphics[width=7in]{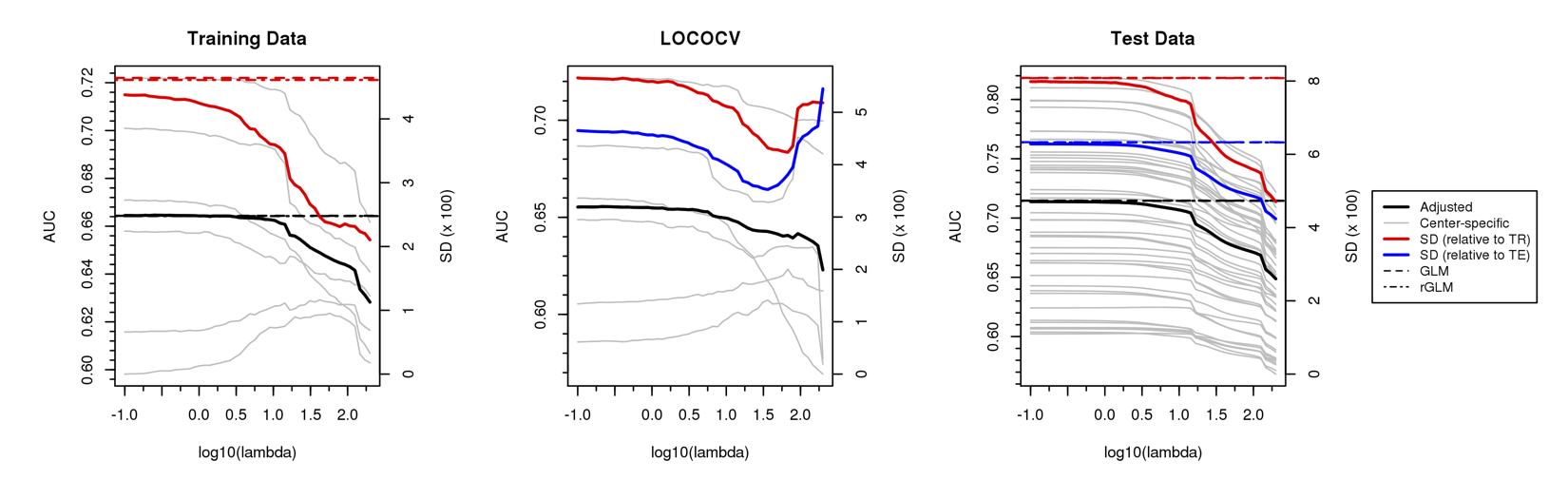}
\captionsetup{labelformat=empty}
\caption{\textbf{Figure 11:} Penalized estimation example 14. The four biomarkers $(X_1, X_2, X_3, X_4)$ were independently normally distributed with mean 0 and center-specific variances, i.e., $(X_i|C) \sim N(0, \sigma_{iC}^2), \;i=1,2,3,4$. The relationships between the biomarkers and $P(D=1)$ were allowed to vary by center and these variations (defined by $(\gamma_{1C}, \gamma_{2C}, \gamma_{3C}, \gamma_{4C})$) differed across the four biomarkers. The outcome $D$ was generated as a Bernoulli random variable with success probability $g(\alpha_0^C + \gamma_{1C}X_1 - \gamma_{2C}X_2 + \gamma_{3C}X_3 - \gamma_{4C}X_4$), where $\alpha_0^C$ is a center-specific intercept and $g(v) = (1+e^{-v/3})^{-1}$ for $v < 0$ and $(1+e^{-3v})^{-1}$ otherwise. In this example, the LOCOCV procedure returns somewhat inconclusive results, making the choice of $\lambda$ less clear.}
\end{figure}

\begin{figure}[ht]
\hspace{-7mm}
\includegraphics[width=7in]{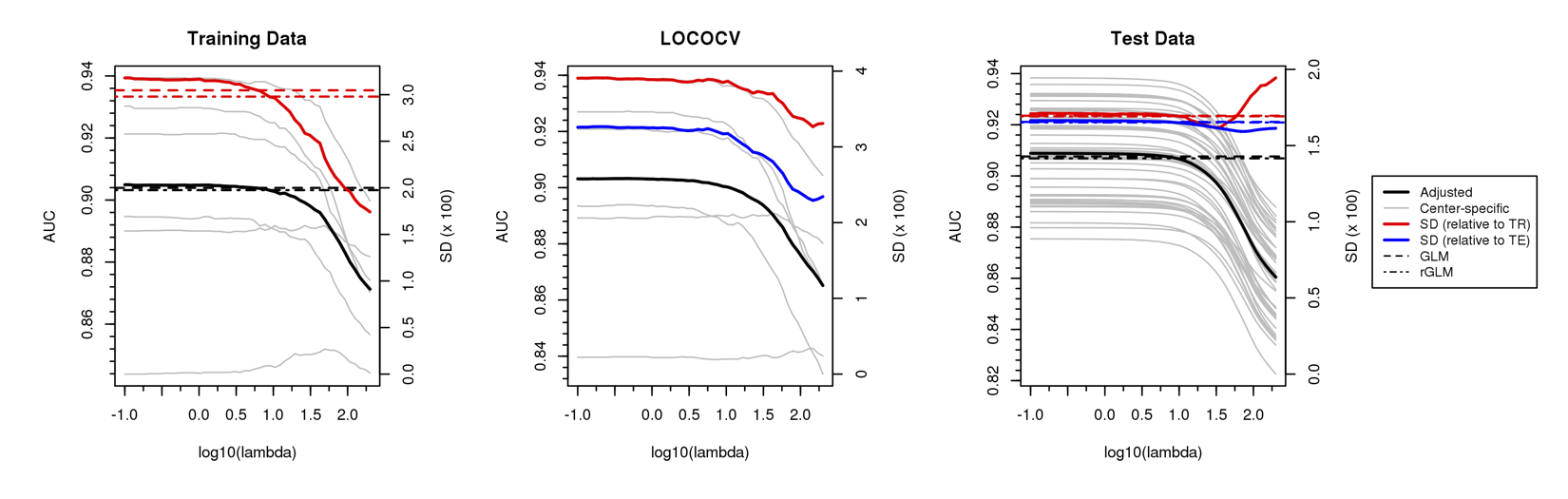}
\captionsetup{labelformat=empty}
\caption{\textbf{Figure 12:} Penalized estimation example 15. The four biomarkers $(X_1, X_2, X_3, X_4)$ were independently normally distributed with mean 0 and center-specific variances, i.e., $(X_i|C) \sim N(0, \sigma_{iC}^2), \;i=1,2,3,4$. The outcome $D$ was generated as a Bernoulli random variable with success probability $g(\alpha_0^C + X_1 - X_2 + X_3 - X_4$), where $\alpha_0^C$ is a center-specific intercept and $g(v) = (1+e^{-v/3})^{-1}$ for $v < 0$ and $(1+e^{-3v})^{-1}$ otherwise. In this example, the LOCOCV procedure is a bit misleading, when compared to the results in the test data.}
\end{figure}

\begin{figure}[ht]
\hspace{-7mm}
\includegraphics[width=7in]{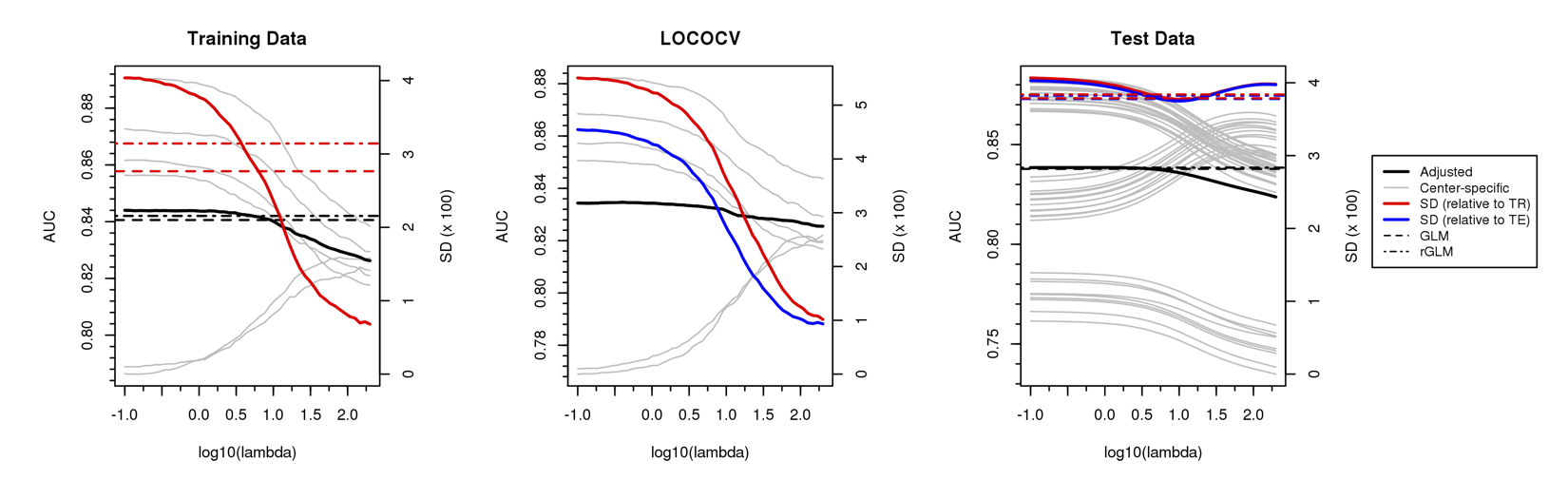}
\captionsetup{labelformat=empty}
\caption{\textbf{Figure 13:} Penalized estimation example 16. The four biomarkers $(X_1, X_2, X_3, X_4)$ were independently normally distributed with mean 0 and center-specific variances, i.e., $(X_i|C) \sim N(0, \sigma_{C}^2), \;i=1,2,3,4$. The relationships between the biomarkers and $P(D=1)$ were allowed to vary by center and these variations (defined by $(\gamma_{1C}, \gamma_{2C}, \gamma_{3C}, \gamma_{4C})$) differed across the four biomarkers. The outcome $D$ was generated as a Bernoulli random variable with success probability $f(\alpha_0^C + \gamma_{1C}X_1 - \gamma_{2C}X_2 + \gamma_{3C}X_3 - \gamma_{4C}X_4$), where $\alpha_0^C$ is a center-specific intercept and $f(v) = (1+e^{-v})^{-1}$. In this example, the LOCOCV procedure is a bit misleading, when compared to the results in test data.}
\end{figure}

\begin{figure}[ht]
\hspace{-7mm}
\includegraphics[width=7in]{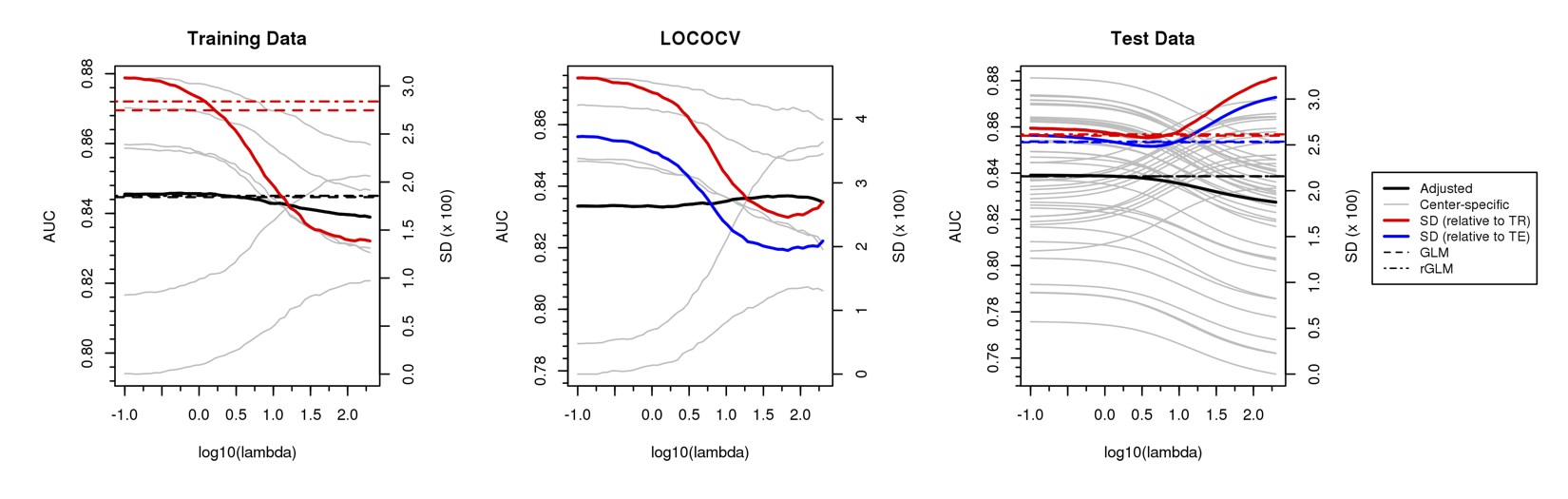}
\captionsetup{labelformat=empty}
\caption{\textbf{Figure 14:} Penalized estimation example 17. The four biomarkers $(X_1, X_2, X_3, X_4)$ were independently normally distributed with mean 0 and center-specific variances, i.e., $(X_i|C) \sim N(0, \sigma_{C}^2), \;i=1,2,3,4$. The relationships between the biomarkers and $P(D=1)$ were allowed to vary by center and these variations (defined by $(\gamma_{1C}, \gamma_{2C}, \gamma_{3C}, \gamma_{4C})$) differed across the four biomarkers. The outcome $D$ was generated as a Bernoulli random variable with success probability $g(\alpha_0^C + \gamma_{1C}X_1 - \gamma_{2C}X_2 + \gamma_{3C}X_3 - \gamma_{4C}X_4$), where $\alpha_0^C$ is a center-specific intercept and $g(v) = (1+e^{-v/3})^{-1}$ for $v < 0$ and $(1+e^{-3v})^{-1}$ otherwise. In this example, the variability in performance in test data increases slightly with increasing $\lambda$, though this is not reflected in the LOCOCV results.}
\end{figure}

\begin{figure}[ht]
\hspace{-7mm}
\includegraphics[width=7in]{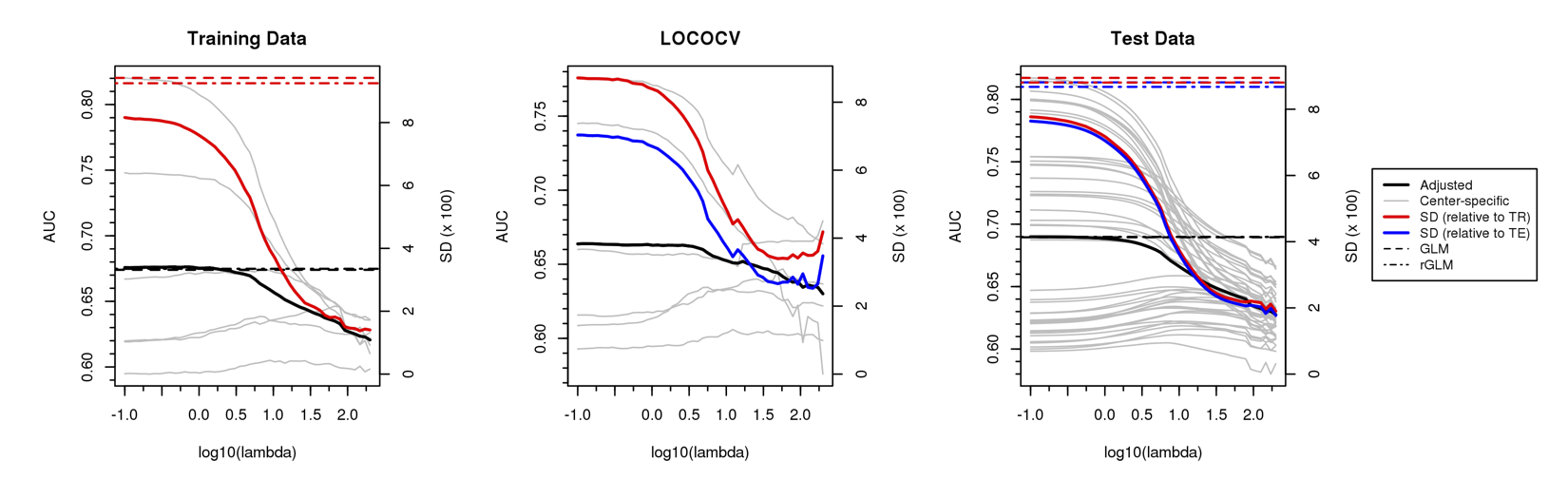}
\captionsetup{labelformat=empty}
\caption{\textbf{Figure 15:} Penalized estimation example 18. The four biomarkers $(X_1, X_2, X_3, X_4)$ were independently normally distributed with mean 0 and center-specific variances, i.e., $(X_i|C) \sim N(0, \sigma_{iC}^2), \;i=1,2,3,4$. The relationships between the biomarkers and $P(D=1)$ were allowed to vary by center and these variations (defined by $(\gamma_{1C}, \gamma_{2C}, \gamma_{3C}, \gamma_{4C})$) differed across the four biomarkers. The outcome $D$ was generated as a Bernoulli random variable with success probability $f(\alpha_0^C + \gamma_{1C}X_1 - \gamma_{2C}X_2 + \gamma_{3C}X_3 - \gamma_{4C}X_4$), where $\alpha_0^C$ is a center-specific intercept and $f(v) = (1+e^{-v})^{-1}$. In this example, the overall performance in test data decreases more rapidly with increasing $\lambda$ than is suggested by the LOCOCV results.}
\end{figure}

\begin{figure}[ht]
\hspace{-7mm}
\includegraphics[width=7in]{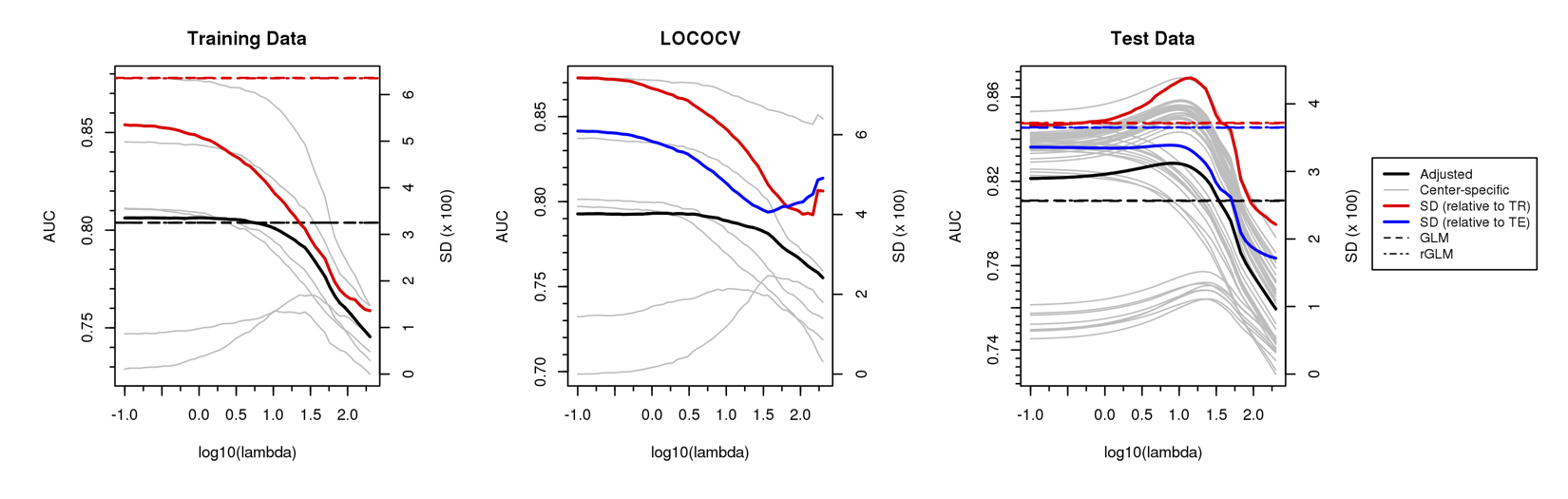}
\captionsetup{labelformat=empty}
\caption{\textbf{Figure 16:} Penalized estimation example 19. The four biomarkers $(X_1, X_2, X_3, X_4)$ were independently normally distributed with mean 0 and center-specific variances, i.e., $(X_i|C) \sim N(0, \sigma_{C}^2), \;i=1,2,3,4$. The relationships between the biomarkers and $P(D=1)$ were allowed to vary by center and these variations (defined by $(\gamma_{1C}, \gamma_{2C}, \gamma_{3C}, \gamma_{4C})$) differed across the four biomarkers. The outcome $D$ was generated as a Bernoulli random variable with success probability $f(\alpha_0^C + \gamma_{1C}X_1 - \gamma_{2C}X_2 + \gamma_{3C}X_3 - \gamma_{4C}X_4$), where $\alpha_0^C$ is a center-specific intercept and $f(v) = (1+e^{-v})^{-1}$. In this example, the relationship between $\lambda$ and variability in performance in the test data is somewhat unusual and is not seen in the LOCOCV results.}
\end{figure}

\clearpage

\bibliographystyle{WileyNJD-AMA}
\bibliography{refs}

\begin{thebibliography}{10}
\providecommand \doibase [0]{http://dx.doi.org/}%

\bibitem{feldstein2009}
Feldstein A, Wieckowska A, Lopez A, Liu YC, Zein N, McCullough A.
  Cytokeratin-18 fragment levels as noninvasive biomarkers for nonalcoholic
  steatohepatitis: a multicenter validation study. {\it Hepatology.}
  2009\string; 50\string: 1072--1078.

\bibitem{degos2010}
Degos F, Perez P, Roche B, et al. Diagnostic accuracy of {F}ibro{S}can and
  comparison to liver fibrosis biomarkers in chronic viral hepatitis: a
  multicenter prospective study (the {FIBROSTIC} study). {\it J Hepatol.}
  2010\string; 53\string: 1013--1021.

\bibitem{nickolas2012}
Nickolas T, Schmidt-Ott K, Canetta P, et al. Diagnostic and prognostic
  stratification in the emergency department using urinary biomarkers of
  nephron damage. {\it J Am Coll Cardiol.} 2012\string; 59\string: 246--255.

\bibitem{parikh2011}
Parikh C, Coca S, Thiessen-Philbrook H, et al. Postoperative biomarkers predict
  acute kidney injury and poor outcomes after adult cardiac surgery. {\it J Am
  Soc Nephrol.} 2011\string; 22\string: 1748--1757.

\bibitem{pepebook}
Pepe M. {\it The Statistical Evaluation of Medical Tests for Classification and
  Prediction}.
\newblock Oxford: Oxford University Press .
\newblock 2003.

\bibitem{janes2008}
Janes H, Pepe M. Adjusting for covariates in studies of diagnostic, screening,
  or prognostic markers: an old concept in a new setting. {\it Am J Epidemiol.}
  2008\string; 168\string: 89--97.

\bibitem{kahan2014}
Kahan B. Accounting for centre-effects in multicentre trials with a binary
  outcome--when, why, and how?. {\it BMC Med Res Methodol.} 2014\string;
  14\string: 20.

\bibitem{janes2009b}
Janes H, Pepe M. Adjusting for covariate effects on classification accuracy
  using the covariate-adjusted receiver operating characteristic curve. {\it
  Biometrika.} 2009\string; 96\string: 371--382.

\bibitem{janes2009}
Janes H, Longton G, Pepe M. Accommodating covariates in receiver operating
  characteristic analysis. {\it Stata J.} 2009\string; 9\string: 17--39.

\bibitem{mcintosh2002}
McIntosh M, Pepe M. Combining several screening tests: optimality of the risk
  score. {\it Biometrics.} 2002\string; 58\string: 657--664.

\bibitem{pepe2006}
Pepe M, Cai T, Longton G. Combining predictors for classification using area
  under the receiver operating characteristic curve. {\it Biometrics.}
  2006\string; 62\string: 221--229.

\bibitem{liu2013}
Liu D, Zhou XH. {ROC} analysis in biomarker combination with covariate
  adjustment. {\it Acad Radiol.} 2013\string; 20\string: 874--882.

\bibitem{schisterman2004}
Schisterman E, Faraggi D, Reiser B. Adjusting the generalized {ROC} curve for
  covariates. {\it Stat Med.} 2004\string; 23\string: 3319--3331.

\bibitem{copas2002}
Copas J, Corbett P. Overestimation of the receiver operating characteristic
  curve for logistic regression. {\it Biometrika.} 2002\string; 89\string:
  315--331.

\bibitem{kerr2015}
Kerr K, Meisner A, Thiessen-Philbrook H, Coca S, Parikh C. {RiGoR}: reporting
  guidelines to address common sources of bias in risk model development. {\it
  Biomark Res.} 2015\string; 3\string: 2.

\bibitem{harrell2001regression}
Harrell F. {\it Regression Modeling Strategies: with Applications to Linear
  Models, Logistic Regression, and Survival Analysis}.
\newblock New York: Springer-Verlag .
\newblock 2001.

\bibitem{bouwmeester2013AJE}
Bouwmeester W, Moons K, Kappen T, et al. Internal validation of risk models in
  clustered data: a comparison of bootstrap schemes. {\it Am J Epidemiol.}
  2013\string; 177\string: 1209--1217.

\bibitem{vanoirbeek2010}
Van~Oirbeek R, Lesaffre E. An application of {H}arrell's {C}-index to {PH}
  frailty models. {\it Stat Med.} 2010\string; 29\string: 3160--3171.

\bibitem{localio2001}
Localio A, Berlin J, Ten~Have T, Kimmel S. Adjustments for center in
  multicenter studies: an overview. {\it Ann Intern Med.} 2001\string;
  135\string: 112--123.

\bibitem{pepe2000}
Pepe M, Thompson M. Combining diagnostic test results to increase accuracy.
  {\it Biostatistics.} 2000\string; 1\string: 123--140.

\bibitem{lin2011}
Lin H, Zhou L, Peng H, Zhou XH. Selection and combination of biomarkers using
  {ROC} method for disease classification and prediction. {\it Can J Stat.}
  2011\string; 39\string: 324--343.

\bibitem{ma2007}
Ma S, Huang J. Combining multiple markers for classification using {ROC}. {\it
  Biometrics.} 2007\string; 63\string: 751--757.

\bibitem{fong2016}
Fong Y, Yin S, Huang Y. Combining biomarkers linearly and nonlinearly for
  classification using the area under the {ROC} curve. {\it Stat Med.}
  2016\string; 35\string: 3792--3809.

\bibitem{meisner2017}
Meisner A, Parikh C, Kerr K. Biomarker combinations for diagnosis and prognosis
  in multicenter studies: principles and methods. {\it Stat Methods Med Res.}
  2019\string; 28\string: 969--985.

\bibitem{hastie2016}
Hastie T, Tibshirani R, Friedman J. {\it The Elements of Statistical Learning}.
\newblock New York: Springer Science \& Business Media .
\newblock 2016.

\bibitem{bianco1996}
Bianco A, Yohai V. Robust estimation in the logistic regression model. In:
  Rieder H. \kern-2pt, ed. {\it Robust Statistics, Data Analysis, and Computer
  Intensive Methods.}New York: Springer-Verlag.  1996 (pp. 17--34).

\bibitem{leborgne2017}
Le~Borgne F, Combescure C, Gillaizeau F, et al. Standardized and weighted
  time-dependent receiver operating characteristic curves to evaluate the
  intrinsic prognostic capacities of a marker by taking into account
  confounding factors. {\it Stat Methods Med Res.} 2018\string; 27\string:
  3397--3410.

\end{thebibliography}

\end{document}